\PassOptionsToPackage{dvipdfmx}{graphicx}
\documentclass[preprint,12pt,authoryear]{elsarticle}

\usepackage{amssymb}
\usepackage{amsthm}
\usepackage{amsmath}

\usepackage{hyperref}

\usepackage[dvipdfmx]{graphicx}
\graphicspath{ {./figures/} }

\usepackage{booktabs}
\usepackage{tabularx}
\newcolumntype{C}[1]{>{\centering\arraybackslash}p{#1}}
\newcolumntype{L}[1]{>{\raggedright\arraybackslash}p{#1}}
\newcolumntype{R}[1]{>{\raggedleft\arraybackslash}p{#1}}

\usepackage{color}

\usepackage{url}

\usepackage[colorinlistoftodos]{todonotes}

\usepackage[utf8]{inputenc}
\usepackage[T1]{fontenc}
\usepackage{lmodern}

\journal{Treatise on Geochemistry, 3rd}

\begin{document}

\begin{frontmatter}



\title{00178. The evolutionary divergence of Mars, Venus and Earth}

\author[inst1]{Keiko Hamano}
\affiliation[inst1]{organization={Earth-Life Science Institute,
Tokyo Institute of Technology},
            addressline={2-12-1-IE-1 Ookayama}, 
            city={Meguro-ku},
            postcode={152-8550}, 
            state={Tokyo},
            country={Japan}}

\author[inst2]{Cedric Gillmann}
\affiliation[inst2]{organization={ETH Z\"{u}rich, Institute of Geophysics}, addressline={Sonneggstrasse 5}, postcode={8092}, city={Z\"{u}rich}, country={Switzerland}}

\author[inst3]{Gregor J. Golabek}
\affiliation[inst3]{organization={Bayerisches Geoinstitut, University of Bayreuth}, addressline={Universit\"{a}tsstrasse 30}, postcode={95440}, city={Bayreuth}, country={Germany}}

\author[inst2]{Diogo Louren\c{c}o}

\author[inst4]{Frances Westall}
\affiliation[inst4]{organization={CNRS-Centre de Biophysique Mol\'{e}culaire}, addressline={Rue Charles Sadron CS 80054}, postcode={45071}, city={Orl\'{e}ans}, country={France}}

\begin{abstract}

Mars, Venus and Earth are expected to have started in a hot molten state. Here, we discuss how these three terrestrial planets diverged in their evolution and what mechanisms could be the cause. 
We discuss that early-on after magma ocean crystallization the mantle/surface redox state and water inventory may already differ considerably, depending on planetary mass and orbital distance from the Sun. During the subsequent internal evolution, the three planets also diverged in terms of their tectonic regime, affecting the long-term planetary evolution via heat flux and outgassing rate, and possibly the physical state of their respective core and the onset and end of a planetary dynamo. We discuss how, throughout the evolution of these rocky planets, the dominant process for atmospheric loss would shift from hydrodynamic escape and impact erosion to non-thermal escape, where small terrestrial planets like Mars are here more vulnerable to volatile loss.

\end{abstract}

\begin{keyword}
atmosphere evolution \sep comparative planetology \sep Earth \sep habitability \sep life \sep magma ocean \sep mantle dynamics \sep Mars \sep planetary evolution \sep Solar System \sep terrestrial planets \sep Venus \sep volatile cycles \sep volcanism
\end{keyword}

\end{frontmatter}


\noindent Key points/objectives box\\

\begin{itemize}
    \item We describe characteristics and processes that played a role for the divergence among the three planets Mars, Venus and Earth.
    \item We discuss how the primitive magma ocean phase could cause the divergence early on.
    \item We investigate long-term interior, surface and atmosphere processes that can enhance those early established differences.
\end{itemize}

\section{Introduction}\label{sec:intro}

Earth, Mars and Venus are all classified as ``terrestrial planets", as opposed to gaseous planets. They have solid surfaces and are composed mostly of silicate rocks and metals with a relatively thin outer atmospheric layer of liquids and gas trapped by gravity. Their building blocks are generally considered to be of similar origin \citep{smrekar2018venus,lammer2018}, coming from the planetary accretion disc, but their respective present-day states are remarkably different. 

The characteristics in the past and present on each planet are described in detail in other chapters, thus here, we briefly compare the three planets. Venus and Earth share similarities in their sizes, masses, and densities, and their orbits are rather close to each other (Table \ref{tab:characteristics}). These characteristics make it possible for Venus to have a comparable structure to Earth, although measurements are currently not precise enough to fully resolve the question \citep[e.g.][]{dumoulin2017}. Comparatively, Mars is smaller and less dense than the two other planets and orbits further away from the Sun.

The most obvious differences between the three planets are readily observable in their atmospheres and surface conditions. Earth has a mild climate with an average temperature of 288 K, and harbors stable liquid water at its surface. It currently has a 1 bar atmosphere mainly composed of nitrogen and oxygen with minor species including noble gases and carbon dioxide. By contrast, Venus and Mars have similar atmosphere compositions dominated by CO$_2$ and N$_2$. SO$_2$ can be observed on Venus (150 ppmv), but very little water (20 ppmv). Mars offers a wide range of species with O$_2$ (0.16\%), CO (0.06\%), H$_2$O and NO. Compared to Earth's atmosphere, Venus has an atmosphere that is about 100 times more massive ($\sim$4.8 $\times~$ 10$^{20}$ kg), while the Martian atmosphere is about 100 times less massive ($\sim$2.5 $\times~$ 10$^{16}$ kg). Surface temperatures also cover a wide range, between Mars' chilly 210 K average (with large variations) and Venus' hellish 737 K. It is interesting to note that, due to its thick atmosphere, Venus boasts a larger atmospheric mass of N$_2$ than the Earth, despite it making up a much smaller percentage of the total surface pressure. Additionally, much of Earth's superficial carbon inventory is stored in carbonates rather than in the planet's atmosphere. This inventory is estimated to be similar to the amount of CO$_2$, present in the atmosphere of Venus \citep{Donahue_Pollack1983,Wedepohl1995,Lecuyer2000,hartmann2012terrestrial}.

\renewcommand{\arraystretch}{1.2}
\begin{table*}[tbp!]
 \caption{Main characteristics of the present Earth, Venus and Mars}
 \label{tab:characteristics}
 \centering
 \small
  \begin{tabularx}{\linewidth}{L{0.5cm} L{0.5cm} C{0.5cm} C{2.8cm} C{2.8cm} C{2.8cm}}
  \toprule
    \multicolumn{2}{l}{} & unit & Earth & Venus & Mars  \\
  \midrule 
  \multicolumn{2}{l}{Radius}&(Earth=1)& 1& 0.949 & 0.532  \\
  \multicolumn{2}{l}{Mass}&(Earth=1)& 1& 0.815 & 0.107  \\
  \multicolumn{2}{l}{Mean density}&(Earth=1)& 1& 0.951 & 0.714   \\
 \multicolumn{2}{l}{Semi-major axis}&au& 1&0.723 & 1.52  \\
  \addlinespace[3mm]
  \multicolumn{2}{l}{Mean surface temperature} & K & 288 & 737 & 210 \\
\multicolumn{2}{l}{Average surface pressure} & bar & 1 & 92 &  6.36$\times10^{-3}$\\
  \multicolumn{2}{l}{Atmospheric composition } & vol.\% &&& \\
  &&&  $\mathrm{N_2(78.08)}$, $\mathrm{O_2(20.95)}$, with noble gases and $\mathrm{CO_2}$ & $\mathrm{CO_2}(96.5)$, $\mathrm{N_2}(3.5)$, and $\mathrm{SO_2}(0.015)$ & $\mathrm{CO_2}(95.1)$, $\mathrm{N_2}(2.59)$, $\mathrm{O_2}(0.16)$, CO(0.06) with $\mathrm{H_2O}$ and NO \\
 \addlinespace[3mm]
  \multicolumn{3}{l}{H$_2$O inventory} & &  \\
   & surface+atmosphere  && 1 ocean &  $<2\times 10^{-4}$~ocean & 20-30~ mGEL (polar caps and permafrost)    \\
  &interior  && 0-6 oceans & - & 100-1000 ppm  \\
  \addlinespace[3mm] 
  \multicolumn{3}{l}{Redox state in the upper mantle}
  & IW+1--5 & - & IW+0--1  \\
  \addlinespace[3mm]
  \multicolumn{3}{l}{Magnetic field}  & &   \\
  & Surface field &T & $5\times10^{-5}$ & $<10^{-8}$ & $10^{-9}-10^{-4}$ (patchy but locally strong)  \\
  & Interpretation && Core dynamo & No dynamo & Remanent crustal field \\
  \addlinespace[3mm] 
  \multicolumn{3}{l}{Tectonic mode} & Plate tectonics & Single plate/ Stagnant lid & Stagnant lid  \\
   \bottomrule
  \end{tabularx}
  \vskip 3pt
\begin{minipage}{0.95\linewidth}
\textit{Data sources}: Radius, mass, mean density, semi-major axis, mean surface temperature, surface pressure and atmospheric composition are from from Planetary Fact Sheet (\url{https://nssdc.gsfc.nasa.gov/planetary/factsheet/planet_table_ratio.html}). H$_2$O inventory from \citet{Peslier+2017SSRv} for Earth and \citet{Scheller+2021Science} for Mars. H$_2$O inventory in Venus's atmosphere was estimated from concentration of water vapor obtained by Venus Express \citep[$\sim$30 ppmv,][]{Marcq+2017}. The unit ``ocean'' 
 represents the Earth's ocean mass, 1.4$\times 10^{21}$ kg, and ``GEL'' represents global equivalent layer. Redox state in the upper mantle from \citet{Righter+2016AM}. For Earth, the redox state shows a large variation, depending on the rock samples (xenoliths, peridotites, MORB, and Kilauea lava samples). Magnetic field informations from \citet{Stevenson2003EPSL}.
\end{minipage}
\end{table*}
\renewcommand{\arraystretch}{1}

Going deeper than surface observation and delving into the interior characteristics of the planets is hampered by the relative paucity of data available beyond Earth. Samples from Mars and in-situ measurements provide some constraints (i.e. water inventory, redox state, isotopic measurements, core size - see Table \ref{tab:characteristics}) that are still unavailable for Venus, which implies that a large part of our understanding of Venus relies on educated guesses based on Earth observations and the consequences of common mechanisms. The present Earth is known for having an oxidized upper mantle, reflected by its oxygen fugacity (a measure of the oxidation state of the mantle, from the ratio of the relative abundance of iron (III) ions to total iron, and depending on pressure and temperature; see also Section \ref{sec:MO}). In Earth's present-day upper mantle, oxygen fugacity relative to the iron-w\"ustite (IW) buffer is estimated to be about IW+1$\sim$IW+5 \citep[][and references therein]{Righter+2016AM,Hirschmann22GCA}, that is between 1 and 5 logarithmic units above the IW buffer. Higher values (up to IW+8) also have been measured in oxidized samples possibly due to processes in the crust or metasomatism. These values are several orders of magnitude higher than the value estimated from FeO content in the upper mantle based on the assumption that mantle material was in equilibrium with Fe metal during core formation ($\sim$ IW-2)\citep[e.g.][]{Frost+2008}. Earth's upper mantle oxygen fugacity appears to have been stable over the last $\approx$2 Gyr \citep{hirschmann2023,stagno2021}. Some heterogeneity is expected depending on the source region of the samples and their evolution \citep{hirschmann2023}. Based on Martian meteorites, the Martian upper mantle seems to have an oxygen fugacity of IW+0$\sim$IW+1\citep[][and references therein]{Righter+2016AM}. This value is also larger than the level set by metal-silicate equilibrium ($\sim$ IW-1) based on FeO content in the Martian mantle, but the difference is much smaller than Earth's. The redox state of the Venusian mantle remains unknown. Calculations based on the equilibration during core formation yield a range of values for oxygen fugacity similar to Earth's \citep{Righter+2016AM}, that would then have probably evolved toward more oxidized values. Venus's surface may be highly oxidized as suggested by thermodynamic calculations and spectroscopic data \citep{Fegley+1997Icarus}.

It is known that Venus shows no signs of a self-generated magnetic field at present-day, even though it has been speculated that remanent magnetisation may still exist in the crust \citep{ORourke2018}. On the contrary, Earth sustains an intrinsic magnetic field at present-day and possibly since the Hadean \citep{tarduno2015}, while Mars shows signs of a past magnetic dynamo in patches of remanently magnetized ancient crustal material \citep{Acuna1999,Plattner-Simons2015}.

Observation of the three planets has also led to hypotheses regarding their internal dynamics. Earth's mantle convection regime is known as plate tectonics \citep[i.e][]{Wessel&Muller2015}, where the strong lithosphere is broken into distinct mobile plates with deformation mostly localized at the plate boundaries and subduction of oceanic plates being the dominant mechanism for driving plate tectonics \citep[e.g.][]{Forsyth-Uyeda1975}. As a result, very little ancient crust still remains at the surface of the Earth. In contrast, Mars exhibits a so-called stagnant-lid regime, where a thick rigid lithosphere is decoupled from the convective mantle and mostly immobile on top of the mantle \citep[e.g.][]{Solomatov1995}. Most of Mars' surface is thus ancient (billions of years old), but shows a specific pattern: a dichotomy between the southern highlands and the northern lowlands, with a $\sim$5 km difference in elevation. Venus is located somewhere in the middle. Its surface shows no sign of (i) global plate tectonics, (ii) localized subduction zones and (iii) a clear topographic dichotomy between oceanic and continental crust as observed on Earth. However, its surface is on average relatively young and signs of deformation have been noted. Several explanations have been proposed, such as episodic activity \citep{turcotte1993} or lithospheric weakening by magma intrusion \citep[the so-called plutonic squishy lid;][]{Rozel2017,Lourenco2018}. The offset between the center of mass and the center of figure of a planet (CM-CF offset) indicates differences in interior dynamics, since they would emerge from substantial large scale density anomalies in the solid planet, possibly due to interior temperature or topography \citep{King2018}. The CM-CF offset has been estimated to be smaller on Venus (280 m, \citet{bindschadler1994venus}) than for Earth (2.1 km, \citet{yoder1995global}) and Mars (3.3 km, \citet{wieczorek2015gravity}). The martian offset has been suggested to be caused by the crustal dichothomy and the Tharsis province \citep{wieczorek2015gravity}. The difference between Earth and Venus is more difficult to explain, with no uniquely identified mechanism that could act as a cause. It has been suggested that the small CM-CF on Venus could be indicative of the lack of substantial downwellings during the last tens to hundreds of million years of the planet evolution and possible lack of a mobile lid regime \citep{King2018}. 

In this chapter, we attempt to provide an overview of recent efforts to clarify what factors and processes make the current state of the three planets so diverse. We focus on physical and chemical processes that occurred after their formation stage, while it should be noted that diversification of terrestrial planets may also originate during their accretion stage \citep[e.g.][]{Jacobson2017,Rubie+2015Icarus}. Section \ref{sec:MO} describes what differences are expected to arise as a result of the formation of deep magma oceans. In Section \ref{sec:postMO}, we address the diversity in tectonic regime of the three planets during the subsequent internal evolution and its consequences on the long-term planetary evolution. How the dominant process for atmospheric loss would shift throughout the evolution of these rocky planets is discussed in Section \ref{sec:atm-loss}. In Section \ref{sec:feat}, we review the unique features in the history of Mars, Venus and Earth that arose from their diverging evolutionary pathways. Finally, we conclude this chapter, with some perspectives in Section \ref{sec:conc}.

\section{Diversity originating during a magma ocean stage}\label{sec:MO}
 Terrestrial planets, especially Earth, likely undergo a single or multiple global melting by giant impacts \citep[e.g.][]{Canup2012,Tucker&Mukhopadhyay2014,Nakajima&Stevenson2015,Lock+2018}. The energy imparted into the planet during a succession of major collision events can be large enough to melt a thick silicate outer layer, possibly down to the core-mantle boundary, producing a deep magma ocean \citep[e.g.][]{Wetherill1976,Drake2000,Nakajima2021}. The planetary characteristics originated during crystallization of a magma ocean would set initial conditions for the subsequent long-term evolution of planetary interior and atmosphere. Here, we focus on two processes that could occur during a magma ocean phase, iron redox disproportionation and appearance of threshold flux of hot steam atmospheres. Recent studies on these two processes highlight the importance of the planetary mass and the orbital distance from the Sun as potential causes of planetary diversity originated during their magma ocean stages, as described in details in this section. We address the proposed mechanisms and predicted characteristics of planets that evolved from a hot molten state.

\subsection{Iron disproportionation in silicate liquid} \label{sec:MO-dispro}
Early planetary mantles could have a redox state which reflects O partitioning among interior reservoirs, affecting the composition of subsequently outgassed volcanic gases and thus long-term climate. One of the mechanisms proposed to affect the redox state in a planetary mantle is iron redox disproportionation,
 \begin{eqnarray}
	3\mathrm{Fe^{2+}} &=& \mathrm{Fe^{0}} + \mathrm{2Fe^{3+}}.
    \label{eq:Fe_dispro}
\end{eqnarray}
The original idea is that crystallization of Fe$^{3+}$-rich solid phases such as bridgmanite (the dominant mineral in the Earth's lower mantle) produces Fe metal by this reaction. Segregation of the produced Fe metal into the core enriches the lower mantle in Fe$^{3+}$. Then, the later mantle convection homogenizes the Fe$^{3+}$-rich lower mantle with the upper mantle, and raises the Fe$^{3+}$ content in the upper mantle gradually \citep[e.g.][]{Frost+2008}. If the same iron disproportionation occurs and produces Fe metal directly in liquid silicate, segregation of Fe metal and homogenization of the remaining Fe$^{3+}$ melt can occur very quickly, possibly increasing the Fe$^{3+}$ abundance in the whole part of the magma ocean \citep{Armstrong+19Sci}.

Under metal-saturated conditions, the partitioning of O between metal and silicate melt determines the oxygen fugacity ($f_\mathrm{O_2}$). 
 
\begin{eqnarray}
	\mathrm{Fe^{met}} + \frac{1}{2}\mathrm{O_2} &=& \mathrm{FeO^{sil}}
    \label{eq:redox_Fe}
\end{eqnarray}
where the superscripts `met' and `sil' indicate a molten metallic phase and a silicate liquid phase, respectively. Ferric iron in silicate melt is also set by the reaction. 
\begin{eqnarray}
	\mathrm{FeO^{sil}} + \frac{1}{4}\mathrm{O_2} &=& \mathrm{FeO_{1.5}^{sil}}.
    \label{eq:redox_Fe3+}
\end{eqnarray}

The molar ratio of Fe$^{3+}$ to total Fe (Fe$^{3+}$/$\Sigma$Fe) in silicate melt increases with the oxygen fugacity, and also varies with pressure, because of the difference in partial molar volumes $\Delta v$ between the ferric and ferrous oxide components  \citep[e.g.][]{Kress&Carmichael91ContribMinPetro,ONeill+06AmMin,Hirschmann12EPSL} \footnote{At an oxygen fugacity fixed relative to a specific oxygen buffer, the pressure effect depends on $\Delta v$ relative to a change in partial molar volumes associated with the buffer reaction \citep{Hirschmann22GCA}.}. This suggests that the Fe$^{3+}$/$\Sigma$Fe ratio in a magma ocean may depend on the planetary mass, since pressures in a magma ocean increase with the planetary mass.  


The pressure dependence of $\Delta v$ and Fe$^{3+}$/$\Sigma$Fe ratios was explored recently by high-pressure experiments \citep{Zhang+17GCA,Armstrong+19Sci,Kuwahara+2023NatGeO} and first principle calculations \citep{Deng+20NatCom}. Fe$^{3+}$/$\Sigma$Fe decreases with pressure up to about 10~GPa owing to the large volume of FeO$_{1.5}$ compared to that of FeO \citep{Kress&Carmichael91ContribMinPetro,ONeill+06AmMin,Zhang+17GCA}. However, this trend reverses at higher pressures because FeO$_{1.5}$ has a larger compressibility than FeO \citep{Armstrong+19Sci,Deng+20NatCom}. Accordingly, Fe$^{3+}$ is more stable at pressures higher than around 10~GPa, possibly up to 90~GPa. These studies show that silicate melt with a higher Fe$^{3+}$/$\Sigma$Fe ratio can be in equilibrium with Fe metal at higher pressure, suggesting that the disproportionation could occur in silicate melt. The increasing stability of Fe$^{3+}$ at high pressures is also reported for basaltic and peridotitic melts based on high pressure experiments by \citet{Kuwahara+2023NatGeO}. Pressures in a magma ocean would be higher for larger planets. Therefore, the enhanced stability of Fe$^{3+}$ suggests that larger planets have a higher abundance of Fe$^{3+}$ in their magma ocean under the same oxygen fugacity and the presence of Fe metal. 


\subsection{Potential correlation between planetary mass and atmospheric composition}\label{sec:MO-Atmtrend}
Fe metal produced by the disproportionation could sink quickly in the magma ocean \citep[e.g.][]{Ichikawa+2010JGR}. If the segregated Fe metal ponds and equilibrates with the magma ocean at its base, then reaction \eqref{eq:redox_Fe} sets the $f_\mathrm{O_2}$ at the base of the magma ocean. Due to the pressure dependence of $\Delta v$ as described above, the $f_\mathrm{O_2}$ in the silicate melt can vary as it ascends along a geotherm in the metal-free magma ocean \citep{Hirschmann12EPSL}, when the Fe$^{3+}$/$\Sigma$Fe ratio is assumed to be constant in a highly turbulent magma ocean. The $f_\mathrm{O_2}$ at the top of a magma ocean is likely higher than that at its base, when the base pressure exceeds roughly 10~GPa \citep{Armstrong+19Sci,Deng+20NatCom,Kuwahara+2023NatGeO}, though some differences exist among these studies. In short, on small planets like the Moon and Mercury, the oxygen fugacity would monotonically decrease from the base of the magma ocean towards the surface \citep{Zhang+17GCA}, while relatively large planets on Earth and Mars the shallow part would be more oxidized than its base \citep{Armstrong+19Sci,Deng+20NatCom}, when the FeO-FeO$_{1.5}$ reaction \eqref{eq:redox_Fe3+} buffers the oxygen fugacity in the magma ocean.

\begin{figure}[tb!]
\centering
\includegraphics[keepaspectratio,width=0.95\linewidth]{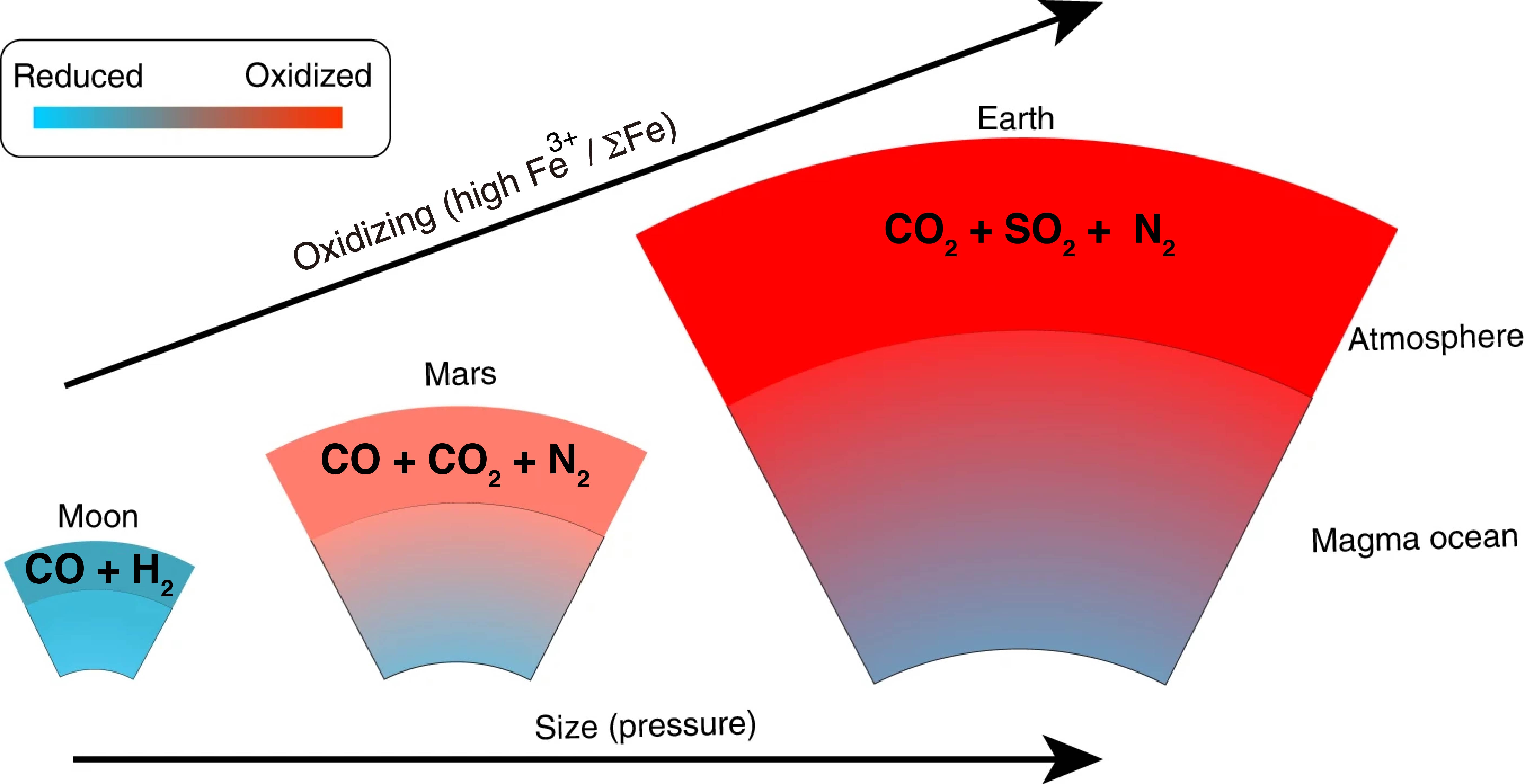}
\caption{The proposed trend in atmospheric compositions formed above a deep magma ocean. The atmospheric composition reflects the surface redox state and the partitioning of C-H-O-N-S elements between a magma ocean and its atmosphere. The oxygen fugacity at the surface is estimated to be IW-2$\sim$-1.5 for the Moon, IW-1$\sim$+1 for Mars, and IW+3$\sim$+5 for Earth based on the assumption of metal-silicate equilibrium at the base of the magma ocean \citep{Deng+20NatCom}. The difference in oxygen fugacity affects the gas speciation and the volatile partitioning. H$_2$O is highly soluble in silicate melt so that its majority is dissolved in the magma ocean. Modified from \citet{Deng+20NatCom} under CC by 4.0 license (\url{http://creativecommons.org/licenses/by/4.0/}), based on numerical results by \citet{Gaillard2022}.}
\label{fig:MO-atmosphere}
\end{figure}

Based on the assumption that metal-silicate equilibration occurred at the base of the magma ocean, these recent findings suggest that a larger planet features more Fe$^{3+}$ in the silicate melt and a more oxidized surface during its deep magma ocean stage. \citet{Deng+20NatCom} proposed a trend between the Fe$^{3+}$ abundance, the surface redox state, the speciation of atmospheres, and the planetary mass (Figure \ref{fig:MO-atmosphere}). As the authors used FeO content in the present upper mantle of Earth, Mars and Moon, the $f_\mathrm{O_2}$ at the base of the magma ocean also differs among the planets, but the overall trend would be similar even if the same $f_\mathrm{O_2}$ is imposed at the base: a larger planet has an oxidizing surface above a Fe$^{3+}$-rich magma ocean.

The oxygen fugacity at the surface controls the speciation of volatile elements in the gas phase \citep{schaefer2017,Gaillard+2021SSRv}. Because volatile solubility in silicate melt strongly depends on gas speciation \citep[see][and references therein]{Lichtenberg+21JGR}, it also affects the partitioning behaviour of volatile elements between a magma ocean and its atmosphere. A gas species having a lower solubility is preferentially outgassed into the atmosphere. The composition of atmospheres formed above a magma ocean also depends on the pressure and temperature, relative abundance of volatile elements, as well as the depth of a magma ocean. \citet{Gaillard2022} modeled the partitioning of C-H-O-N-S elements between a deep magma ocean and its atmosphere at 1,500 ${}^\circ$C using volatile inventories on the present Earth. According to the scenario proposed by \citet{Deng+20NatCom} and the numerical results by \citet{Gaillard2022}, large planets like Earth could have an oxidizing atmosphere consisting of CO$_2$, SO$_2$, and N$_2$ during its deep magma ocean stage (Figure \ref{fig:MO-atmosphere}). Although we have little information about the Venusian mantle, Venus could be expected to have had an oxidizing atmosphere by analogy, if Venus experienced metal-silicate equilibration at high pressures as well. Intermediate-sized planets like Mars might have a weakly reducing atmosphere consisting of CO, CO$_2$, and N$_2$. And if planets as small as the Moon formed with C-H-O-N-S whose relative abundance is as the same as that of the Earth, the shallow magma ocean might produce a highly reducing atmosphere consisting of CO and H$_2$.

The oxidation mechanism involving the high stability of Fe$^{3+}$ in silicate melt at high pressures seems a promising way to oxidize a large part of the silicate portion of planets rapidly. On the other hand, the Fe$^{3+}$ abundance in the magma ocean estimated for each planet relies on the simplified assumption of perfect equilibration between silicate melt and metal at the base of the magma ocean. For Earth, the estimated Fe$^{3+}$/$\Sigma$Fe ratios show significant differences among studies and do not necessarily coincide with the observations (Section \ref{sec:earth}). The cause for the differences between the predicted and observed Fe$^{3+}$/$\Sigma$Fe ratios in the mantle requires further investigation.

The solubility of volatile elements also depends on the chemical composition of silicate melts and temperature, as well as cross-interaction among dissolved volatiles, and much remains to be determined especially for extreme conditions relevant to deep magma oceans (e.g. ultramafic compositions and high temperatures). Under more reducing conditions, the formation and flotation of graphite also would be an important factor on the C budget by controlling the oxygen fugacity and limiting the amount of outgassed C \citep{Keppler&Golabek2019}. On planets like Earth, another C-bearing phase, namely diamonds, possibly forms due to the reducing condition at greater depths \citep{Armstrong+19Sci}. Keeping C as diamonds in the interior could cause the atmosphere and surface to be C-depleted \citep{Dasgupta2013RevMinGeochm}. The precipitation and stability of diamonds in a deep magma ocean remains to be investigated.

Note that the atmospheric compositions predicted for a deep magma ocean stage in Figure \ref{fig:MO-atmosphere} would be different from those finally remaining at the end of the magma ocean stage. Later outgassing processes along with magma ocean crystallization can alter the mass and composition of atmospheres (see the next section). The redox state in a magma ocean can also vary due to changes in Fe$^{3+}$/$\Sigma$Fe as crystallization proceeds \citep[e.g.][]{Boujibar+2016,Davis+2021}. According to recent modeling results by \citet{Maurice+2023PSJ}, the Fe$^{3+}$/$\Sigma$Fe ratio and the oxygen fugacity in the crystallizing magma ocean can increase by several orders of magnitude throughout magma ocean crystallization. Although this would not change the total amount of Fe$^{3+}$ in the mantle, the significant oxidation in the shallow part could cause a transition from a reducing atmosphere towards an oxidizing atmosphere during the solidification process.

\subsection{Runaway threshold as a lower limit of thermal radiation from hot steam atmospheres}\label{sec:MO-runaway}

Following core formation, a magma ocean crystallizes by convective heat transport. At high degrees of melting, ultramafic melts are known to have a very low viscosity \citep{Liebske2005,Posner2018}. Rayleigh numbers of a planetary-scale magma ocean are extremely high, ranging from $10^{20}$ to $10^{30}$ \citep[e.g.][]{Solomatov07Treatise}, thus the convection in a deep magma ocean is supposed to be highly turbulent. At the early stage of solidification, volatile species close to the surface are expected to dissolve in the surface magma and to be transported downwards and homogenized by the vigorous convection. As solidification proceeds, dissolved volatiles would be released towards the surface, forming an overlying atmosphere. The outgassed atmosphere insulates the surface and slows the cooling of a magma ocean. Recently, such an interplay between a magma ocean and planetary atmosphere has been modeled and investigated numerically by several groups \citep[e.g.][]{Elkins-Tanton08EPSL,Hamano+13Nature,Hamano+15,Lebrun+13JGR,Schaefer+2016,Salvador+17JGR,Katyal_2019,Bower+19A&A,Bower+22PlanetSciJ,Lichtenberg+21JGR}. According to theoretical studies, the effectiveness of thermal blanketing depends on the mass and composition of overlying atmospheres. In particular, if a steam-dominated atmosphere forms during magma ocean crystallization, terrestrial planets could experience different evolutionary paths despite the same initial state, where the path is being controlled by the orbital distance from the Sun \citep{Hamano+13Nature,Lebrun+13JGR,Salvador+17JGR}.

\begin{figure}[tb!]
\centering
\includegraphics[keepaspectratio,width=0.85\linewidth]{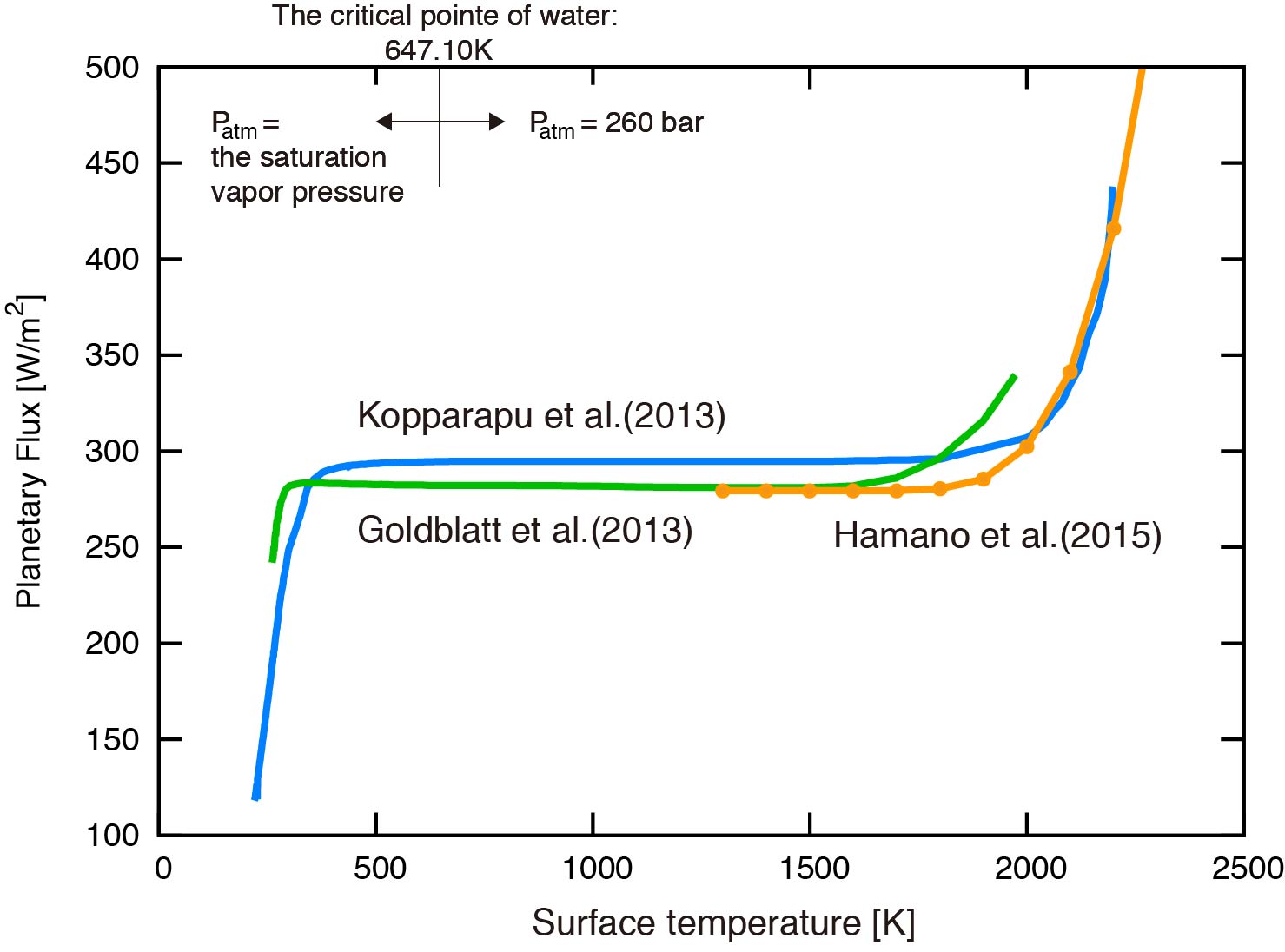}
\caption{Thermal radiation from steam atmospheres. For these radiative transfer calculations, the atmospheric pressure ($P_\mathrm{atm}$) is given by the saturation vapor pressure of water at surface temperature below the critical point of water, and set to 260~bar above it. As the surface temperature increases, the thermal radiation increases and then levels off at about 280~W/m$^2$ (radiation limit or runaway threshold), thus the runaway threshold is an upper limit in thermal radiation from water-saturated steam atmospheres. Above the critical point of water, the thermal radiation gradually starts to increase from the radiation limit with increasing surface temperature, owing to thermal emissions from the hot surface and the hot and dry (unsaturated) lower atmosphere. Under these hot surface conditions, the threshold flux corresponds to a lower limit in the thermal radiation from steam atmospheres. Modified from \citet{Hamano+15}.}
\label{fig:threshold}
\end{figure}

Why is the formation of a steam-dominated atmosphere important for the possible divergent evolution of terrestrial planets? This is because it can limit the thermal radiation from hot molten planets to a certain value. Thermal radiation from steam atmospheres has been extensively explored especially regarding conditions of the formation of liquid water ocean \citep{Komabayasi67JMetSocJapan,Ingersoll69JAtmSci,Rasool&DeBergh70Nature,Pollack71Icarus, Abe&Matsui88JAtmosSci,Kasting88Icarus,Goldblatt+13NatGeo,Kopparapu+13ApJ}. Radiative transfer calculations show that moist (water-saturated) atmospheres have an upper limit for the outgoing infrared flux that can be emitted from its top. If any energy input exceeds this limit, the surface temperature increases until a new radiative equilibrium state is reached and all liquid water completely evaporates from the surface. This is called a runaway greenhouse state.

The physical mechanism behind the radiation limit was elucidated by 
\citet{Nakajima+1992JAS} as follows.
The thermal radiation from the top of the atmosphere $F_\mathrm{pl}$ is given by
\begin{eqnarray}
    F_\mathrm{pl} &=& \int F_\lambda~d\lambda
\end{eqnarray}
and
\begin{eqnarray}
    F_\lambda &=& \int_0^{\tau_{\lambda,s}} \pi B_\lambda(\tau') \exp{[-\tau']} d\tau' + \pi B_\lambda(\tau_{\lambda,s})\exp{[-\tau_{\lambda,s}]}
    \label{eq:Flambda}
\end{eqnarray}
where $\tau$ is the effective optical depth at the top of the atmosphere, and $F_\lambda$ and $B_\lambda$ is the spectral irradiance and the Planck function emitted at wavelength $\lambda$. The subscript $s$ represents the planetary surface. The first term at the right hand side of eq.(\ref{eq:Flambda}) represents contributions of thermal emission from atmospheric layers and the second term is the emission from the planetary surface. The thermal radiation is most sensitive to the temperature around $\tau \sim 1$.

At surface temperatures lower than the critical point of water, the amount of H$_2$O is roughly governed by the saturation water vapor pressure. At low surface temperatures, the atmosphere is optically thin so the planetary radiation comes mainly from the surface. 
As the surface temperature increases, more water evaporates from the surface so the atmosphere becomes dominated by water vapor. The emission from the surface attenuates within the opaque atmosphere, while the infrared thermal radiation at the top of the atmosphere mainly comes from a moist convective region (troposphere), in which the optical depth becomes $\sim$ 1.

Assuming that the steam-rich atmosphere is saturated with water vapor, its thermal structure (a pressure-temperature relation) in the moist convective region approaches asymptotically the saturation vapor pressure curve of water vapor. The pressure level at which the optical depth is equal to unity also lies on the saturated vapor pressure curve. Consequently, the thermal radiation from the top of the atmosphere also approaches a certain value (radiation limit or runaway threshold) as the surface temperature increases (Figure \ref{fig:threshold}).

If the absorbed solar flux exceeds the critical flux, the surface temperature increases until the outgoing thermal flux balances with the net solar insolation. Above the critical point of water, the thermal radiation starts to increase again as surface temperature increases (Figure \ref{fig:threshold}). This is because the thermal emissions from the hot surface and the hot and water-unsaturated lower atmosphere escape through atmospheric windows, thus increasing both the first and second terms at the right hand side of Eq.\eqref{eq:Flambda} \citep{Goldblatt+13NatGeo, Hamano+15}. When the mass of the steam atmosphere increases, the thermal radiation starts to increase above the radiation limit, but never falls below the runaway threshold. In short, under hot surface conditions, like those during magma ocean crystallization, the radiation limit corresponds to a lower limit for the outgoing thermal radiation of steam atmospheres. It can be speculated that the presence of such a limit affects the energy budget of solidifying terrestrial planets, depending on whether the net incoming flux from the young Sun exceeds the lower limit or not. In a runaway greenhouse state, the surface temperature could exceed the solidus temperature of rocks \citep[e.g.][]{Kasting88Icarus}. A 30~bar steam atmosphere is sufficient for surface temperature to exceed the solidus temperature of peridotite rocks with the Earth's gravity \citep[e.g.][]{Hamano20OUP}.

\subsection{Two distinctive cooling paths depending on the distance from the Sun}\label{sec:MO-cooling}
The effect of the radiation limit was first incorporated into a coupled model of a magma ocean and its overlying atmosphere by \citet{Hamano+13Nature} and \citet{Lebrun+13JGR}. In both studies, the authors calculated the thermal evolution of molten terrestrial planets along with the formation of outgassed atmospheres. \citet{Hamano+13Nature} considered a simple gray model of a pure H$_2$O atmosphere overlying a magma ocean and also incorporated loss of water associated with hydrodynamic escape of hydrogen in their model. In the \citet{Lebrun+13JGR} model, the escape process of atmospheres was neglected, while radiative transfer of H$_2$O--CO$_2$(--N$_2$) atmospheres was treated in a more realistic way with non-gray gas opacities.

\begin{figure}[tb]
\centering
\includegraphics[keepaspectratio,width=0.95\linewidth]{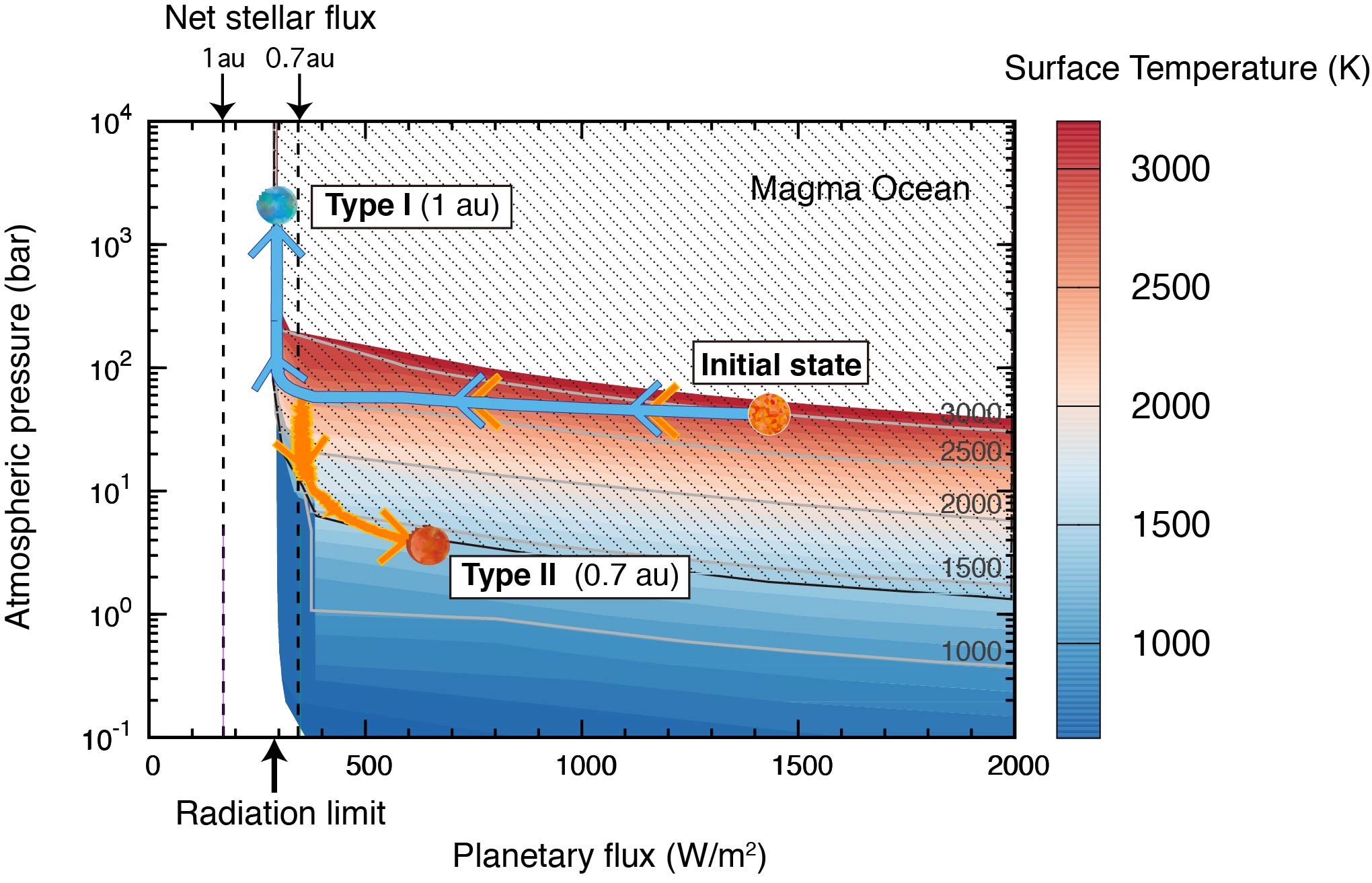}
\caption{Two types of evolutionary paths during magma ocean solidification. Both paths start at the same initial condition at 3,000~K. On a planet at 1~au (``Type I"), the atmosphere monotonically grows during cooling. In contrast, a planet at 0.7~au (``Type II") loses its atmosphere after the planetary radiation reaches the net stellar radiation at the orbit. Calculation results from \citet{Hamano+13Nature}.}
\label{fig:MO-cooling-paths}
\end{figure}

\begin{figure}[tb!]
\centering
\includegraphics[keepaspectratio,width=0.8\linewidth]{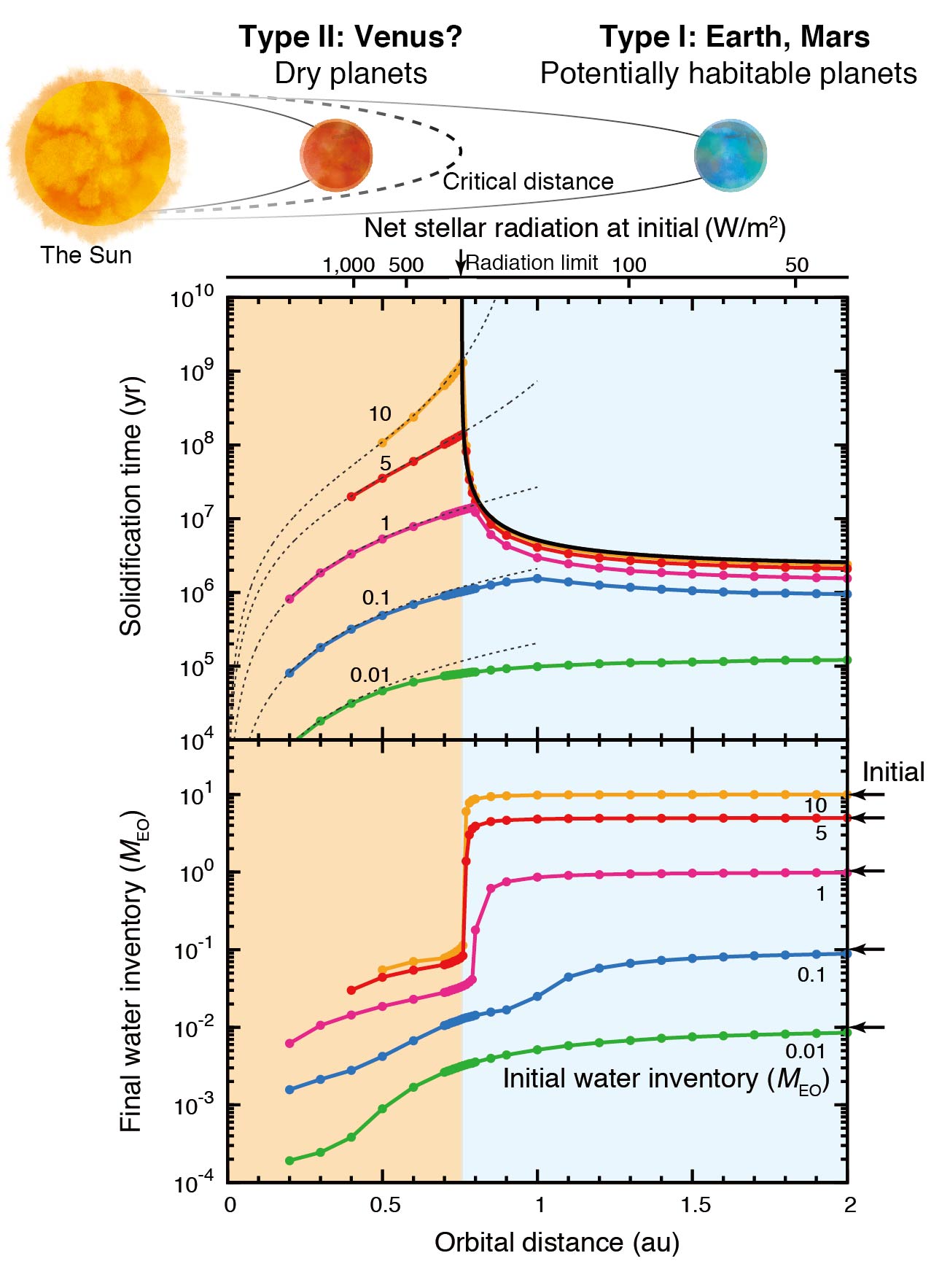}
\caption{Evolutionary dichotomy for terrestrial planets. Beyond the critical distance (see text for its definition), an Earth-sized planet solidifies on the order of 10$^6$ years or less, keeping its water (Type I). In contrast, inside the critical distance, a magma ocean could last longer for a larger inventory of water, and moreover, its solidification time is well approximated by the time required for total loss of the primordial water (Type II), indicated with thin dotted curves in the top panel. In short, a Type II planet is desiccated during the slow solidification process. Modified from \citet{Hamano+13Nature}}
\label{fig:MO-cooling-tsol}
\end{figure}

\citet{Hamano+13Nature} and \citet{Lebrun+13JGR} assumed that volatiles, which are initially dissolved in a deep magma ocean are exsolved at the surface as solidification proceeds. They evaluated the thermal blanketing effect by an overlying atmosphere, whose mass and composition also evolve by volatile outgassing from the interior and atmospheric escape to space. The results for a pure steam atmosphere are shown in Figure \ref{fig:MO-cooling-paths}. The evolutionary paths of solidifying Earth-sized planets formed at 1 au (Type I) and 0.7 au (Type II) are imposed on results of radiative transfer calculations. Based on the assumption of dissolution equilibrium between the magma ocean and the overlying atmosphere, both planets start out with an about 40~bar steam atmosphere with a total H$_2$O inventory of 5 Earth's ocean masses. At a surface temperature of 3,000~K, the initial thermal radiation at the top of the atmosphere amounts to approximately 1,500~W/m$^2$.

On the Type I planet, the absorbed solar flux is lower than the radiation limit. As the thermal radiation drops due to a decrease in surface temperature, the planet moves in this diagram from right to left until it reaches the threshold flux. The subsequent cooling heat flux is given by a flux difference between the radiation limit and the absorbed solar flux, and is high enough for the outgassing of water to overwhelm the escape. Therefore, the curve for the Type I planet bends upward along the radiation limit due to a net growth of the steam atmosphere. The resulting crystallization timescale is on the order of 1~Myr or less. Due to the short duration, most water remains at the end of the crystallization of the magma ocean (Figure \ref{fig:MO-cooling-tsol}). The remaining dense steam atmosphere condenses and forms an ocean immediately after the end of the magma ocean phase \citep{Lebrun+13JGR}.

In contrast to the Type I planet, the solar radiation absorbed by the Type II planet exceeds the radiation limit due to the proximity to the young Sun. In this diagram, the Type II planet starts at the same position as the Type I and moves from right to left as well. However, it reaches the absorbed solar flux, prior to the threshold flux. The Type II planets cannot move further to the left, otherwise the absorbed flux exceeds the outgoing thermal radiation, thus leading to an increase in the surface temperature and the thermal radiation. As all surface H$_2$O is present as water vapor that can be rapidly lost by photodissociation and followed by hydrogen escape to space. The net loss of the steam atmosphere allows the planet to lose its heat by making the atmosphere optically thin enough to emit the thermal radiation slightly larger than the absorbed solar flux. This is why the curve for the Type II planet sharply bends downward approximately along the value of the insolation. As a result, the Type II planet is desiccated and much less water is left on the surface when it completely solidifies (Figure \ref{fig:MO-cooling-tsol}). In the case that internal heat flux from a magma ocean is extremely low, surface temperature may also decrease due to structural changes in steam atmospheres, possibly leading to a shorter-lived magma ocean \citep{selsis2023}. This possibility remains to be examined further, with a coupled magma-ocean-atmosphere model, which includes outgassing processes.

A similar dichotomy was also reported for CO$_2$-H$_2$O atmospheres by \citet{Lebrun+13JGR}, \citet{Salvador+17JGR} and \citet{Massol2023}. A critical condition that separates these two evolutionary paths is given by the balance of absorbed solar flux and the critical threshold for a runaway greenhouse. This condition can be expressed as a critical orbital distance $a_\mathrm{cr}$ from the Sun\footnote{The expression for $a_\mathrm{cr}$ is re-scaled with the critical flux for Earth-sized planets obtained by \citet{Goldblatt+13NatGeo} and \citet{Kopparapu+13ApJ} (about 280 W/m$^2$).} \citep{Hamano+13Nature},
\begin{eqnarray}
    a_\mathrm{cr} 
    &\approx& 0.77~\left( \frac{F_\mathrm{lim}}{280~\mathrm{W/m^2}} \right)^{-1/2} \left(\frac{S}{0.7~S_\mathrm{sun}}\right)^{1/2}
    \left(\frac{1-\alpha_\mathrm{pl}}{1-0.3} \right) ~\mathrm{[au]}
    \label{eq:acr}
\end{eqnarray}
where $S$ and $S_\mathrm{sun}$ are the past and present solar constants, $\alpha_\mathrm{pl}$ is the planetary albedo, and $F_\mathrm{lim}$ is the radiation limit, which depends on the planetary gravity. The threshold value obtained from non-gray cloud-free radiative transfer calculations is about 280~W/m$^2$ for the Earth's gravity and about 250~W/m$^2$ for the Mars' gravity \citep{Goldblatt+13NatGeo,Kopparapu+14ApJ}. Therefore, the critical distance for Mars-sized planets is about 6~\% larger than that for Earth-sized planets. According to the value of the critical distance (Eq. \ref{eq:acr}), Earth and Mars are categorized as Type I. If they had acquired abundant water during formation, both planets would have formed oceans soon after their solidification period on timescales of $\sim$10$^6$ years or less. By contrast, it is not obvious which planetary type Venus belongs to. Three different scenarios that Venus might have followed are discussed in Section \ref{sec:venus}.

The critical distance is identical to the inner edge of the habitable zone, which is set by the onset of a runaway greenhouse state (the so-called runaway greenhouse limit). According to recent 3D simulations, the onset condition for a runaway greenhouse also depends on the initial state of the planetary atmosphere due to different cloud feedbacks \citep{Way+2020JGR,Turbet+21Nature}. If Venus started from an extremely hot condition (magma ocean) like Earth, the Type II scenario seems more probable \citep{Turbet+21Nature}. In our solar system $a_\mathrm{cr}$ lies in the middle of the formation region of the terrestrial planets. This is simply because the critical flux of steam atmospheres is comparable to the typical net insolation at these orbits. The concept of a runaway greenhouse itself is applicable to atmospheres dominated by other condensable gas species such as CO$_2$ and CH$_4$. However, the critical flux estimated for these gas species is much smaller than that for water vapor \citep{Pierrehumbert2010,McKay+99Icarus}, making the location of the runaway greenhouse limit for other gas species irrelevant for the terrestrial planet region.

In reality, planets would acquire not only water vapor, but also other gas species during accretion. Many modeling studies have treated C-O-H atmospheres, especially oxidizing atmospheres consisting of CO$_2$--H$_2$O, in a more general case \citep{Elkins-Tanton08EPSL,Lebrun+13JGR,Salvador+17JGR,Bower+19A&A,Bower+22PlanetSciJ}. \citet{Sossi+SciAdv} found that early atmospheres are depleted in H-bearing components because most H is dissolved in the magma ocean as H$_2$O and the main component for the Earth's earliest atmosphere could be CO. The similar feature of the H-depleted atmosphere and the hydrous magma ocean is seen in calculations of the C-H-O partitioning \citep{Bower+22PlanetSciJ} and the C-H-O-N-S partitioning \citep{Gaillard2022} between a deep magma ocean and its atmosphere in a wide range of surface oxygen fugacity. \citet{Gaillard2022} also found that, under highly reducing conditions with oxygen fugacity less than IW-4, most H is present in the atmosphere as H$_2$ and the atmosphere is C-depleted owing to the formation of graphite at the surface. At a fixed $f_\mathrm{O_2}$, \citet{Bower+22PlanetSciJ} also calculated the change in the mass and composition of the atmospheres by outgassing processes. They showed that later outgassing of H$_2$O and CO$_2$, as well as a decrease in surface temperature, increases the amount of H-bearing species such as H$_2$O and H$_2$, and the CO$_2$/CO ratio in the atmosphere. 

The formation of H$_2$O-dominated atmospheres would depend on the outgassing efficiency at the late stage of solidification, as well as the volatile inventories and the evolution of the surface redox state. As long as the melt fraction in the surface magma is high enough for it to behave like a liquid, water dissolved in the magma ocean likely outgases fast enough to keep the system nearly at dissolution equilibrium \citep{Pahlevan+19EPSL,Hamano+13Nature}. The efficient outgassing would occur by coalescence and ascent of bubbles in a low viscosity magma flow \citep{Massol+16SSRv} or the diffusion of water molecules through a surface thermal boundary layer, which can be as thin as on the order of centimeters with a low liquid-like magma viscosity \citep{Solomatov07Treatise}. On the other hand, when the melt fraction drops along with crystallization so that rheological transition from liquid-like to solid-like behavior occurs, the outgassing rate of water may decrease. At this stage, a highly viscous boundary layer would start to grow at the top of the magma ocean. Although melt percolation seems to transport volatiles efficiently upwards\cite[e.g.][]{Lebrun+13JGR}, its velocity decrease as the melt fraction decreases. Some of the volatiles including water might be trapped in the highly viscous crystal mush under the thick rigid surface layer. \citet{Bower+22PlanetSciJ} suggest that the outgassing efficiency at late stage of a magma ocean period would depend on the style of crystallization, i.e. fractional crystallization or equilibrium crystallization. The outgassing efficiency may also depend on the vigor of convection and the water abundance in the magma ocean \citep{Salvador&Samuel2023}. Even if the oxygen fugacity is extremely low at an early stage so that most of H is present as H$_2$ in the atmosphere, subsequent fractionation of Fe$^{2+}$ and Fe$^{3+}$ during the crystallization process may gradually oxidize the surface magma ocean, converting H$_2$ into H$_2$O, possibly leading to the transition to a steam-dominated atmosphere \citep{Maurice+2023PSJ}.

\section{Post-magma ocean internal evolution}\label{sec:postMO}

The long-term evolution of a planet's interior profoundly impacts surface conditions through the release of volatile species from the mantle into the atmosphere. Here, we examine how terrestrial planets in the Solar System could diverge during their long-term post-magma ocean internal evolution. Furthermore, we discuss the current knowledge regarding the differences in interior evolution for Mars, Earth and Venus.
The interior evolution of terrestrial planets is governed by heat transport in the mantle and its loss at the surface. Heat is generated by the decay of radioactive elements incorporated into the planet during formation, plus any residual accretion/differentiation heat. This determines the dynamics of the planet's mantle and temperature field, but also partial melting of mantle material and magmatism/volcanism, mantle structure, topography, and even possible magnetic field generation \citep[e.g.][]{schubert2001}. Heat can be transported through convection, which is the more efficient mechanism, or through conduction.

\subsection{Convection regimes}
\begin{figure}[tb!]
\centering
\includegraphics[keepaspectratio,width=0.75\linewidth]{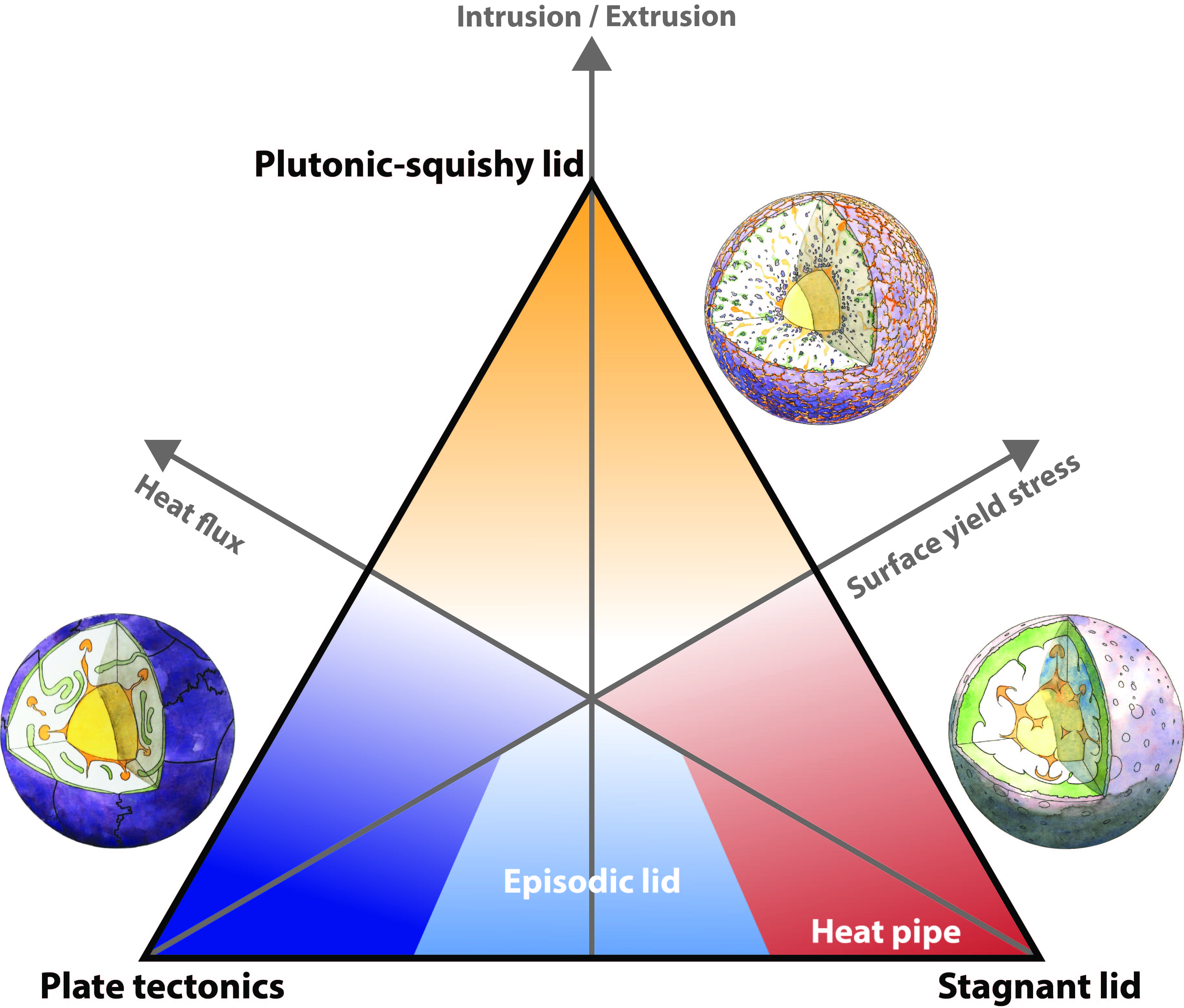}
\caption{The three main end-member convection regimes: Plate tectonics, Stagnant Lid and Plutonic-squishy lid. Two additional specific cases are also highlighted: episodic lid, which is the transition between plate tectonics/mobile lid and the stagnant lid end member, and heat pipe, a stagnant lid with a large volcanic production (i.e., eruption). Schematic views of the interior dynamics of the three main regimes are indicated as cutaway illustrations of planetary interior structures, next to their respective domains. Regimes are distinguished by possibly observable characteristics of the planet. These axes should not be considered as the causes of the regimes depicted and are not fully independent from each other. Surface heat flux, the most readily available observable in the solar system, depends on many factors, such as size, mass, age and composition of the planet. The intrusion to extrusion ratio defines how much of the melt produced in the upper mantle reaches the surface. It is not a direct proxy of total volcanism and is likely to vary spatially. It could be a function of the surface temperature, which is a more direct observable, but can change over time. Surface yield stress is the convective stress caused by mantle convection required to break the lithosphere. It varies with the age of the planet and probably surface temperature. The presence of surface water and other fluids, and processes such as erosion, are also known to facilitate subduction and therefore lower the surface yield stress, which could introduce additional points of divergence between Mars, Venus and Earth. Lithospheric thickness could be considered as a first order observable for this parameter, although it is also influenced, to a lesser extent, by the intrusion to extrusion ratio. Convection regime figures are modified from \citet{Lourenco-Rozel2023}.}
\label{fig:regimes}
\end{figure}

A convection (or tectonic) regime is the surface expression of interior dynamics. In general, various primary convection dynamic regimes are considered: mobile lid (with plate tectonics being a specific subset), stagnant lid, heat pipe, and plutonic-squishy lid (see Figure \ref{fig:regimes}).

The mobile lid regime is characterized by its surface mobility, as the lithosphere is part of the convecting cell and surface velocities are equal to or greater than those of the underlying mantle. In the specific case of plate tectonics, the surface is broken into a small number of plates exhibiting differing relative motion, with deformation highly concentrated along the narrow boundaries between these plates. Plate tectonics is further characterized by long-term continuous subduction processes \citep[e.g.][]{tackley00c, stein04, Rolf2022}.\\ 

In a stagnant lid, a rigid lithosphere shows no lateral mobility and little deformation \citep{Nataf:1982gc,CHRISTENSEN:1984bq,Solomatov1995}. The lid sits on top of a convective mantle; the hot mantle and cold lithosphere are decoupled. Stagnant lid naturally occurs when planetary lithospheres cannot be broken or mobilized by mantle convection, either because the lithosphere is too strong or the convective stresses are too low. Therefore, planets with a thick lithosphere or sluggish convection, for example, would fall into this regime \citep[e.g.][]{Sleep2000,o2016,stern2018,Rolf2022}. As a result, heat flux and volcanic production are expected to be generally lower on stagnant lid than on mobile lid planets. For small or old planets, stagnant lid appears to be the final regime as geological activity dies down and the end-member toward which all terrestrial planets ultimately evolve as they cool down. 

However, for larger, younger planets, the stagnant lid can insulate the mantle from the surface because heat transport by conduction through the lid is inefficient. Reducing the heat loss from the mantle leads to increasing interior temperatures that can cause substantial partial melting. A subset of the stagnant lid is a direct result of this scenario and is called the heat pipe regime, where heat is transported to the surface through erupting silicate melt \citep{moore13,moore2017}. This subset of the stagnant lid regime has been suggested for young terrestrial planets and, at present-day, for Io, one of Jupiter's moons \citep{OReillyDavies}. In such cases, volcanic production and heat transport can be as high as in the plate tectonics regime.

Plutonic-squishy lid has been suggested recently as an alternative to the previous scenarios, where most of the melt produced by the planet is intruded, and remains within the lithosphere rather than reaching the surface. This leads to a warm ductile lithosphere that prevents plate tectonics \citep{Rozel2017,Lourenco2018,Lourenco2020}. However, the surface can be moderately mobile (more so than in stagnant lid but less than in mobile lid regime), but surface velocities are lower than the convective mantle's. Deformation is more localized than in stagnant lid but less than in a plate tectonics, with generally small tectonic units. Heat is extracted from the mantle more efficiently than in the stagnant lid regime. Isolated subduction by lithospheric dripping is possible. Volcanic production would be relatively low due to the high fraction of intruded melt, but the surface heat flux could be close to that of the mobile lid regime \citep{Lourenco2018}.\\

The regimes discussed above are possible end-members for what is likely closer to a continuum of interior dynamics. Real-life is more complicated than these ideal configurations and spatial and temporal variations are expected for each given planet. It is also likely that terrestrial planets can switch from one regime to the other during their evolution \citep{sleep2000evolution}. An episodic lid regime, for example, could correspond to a scenario with short occurrences of mobile lid separated by long periods of stagnant lid \citep[e.g.][]{armann2012} or the other way around \citep{lenardic2012notion}. Finally, it is likely that transition periods between regimes could last hundreds of millions to billions of years \citep[e.g.][]{weller2020,Rolf2022}.\\

\begin{figure}[tb!]
\centering
\vspace*{-3.5cm}
\includegraphics[keepaspectratio,width=0.7\linewidth]{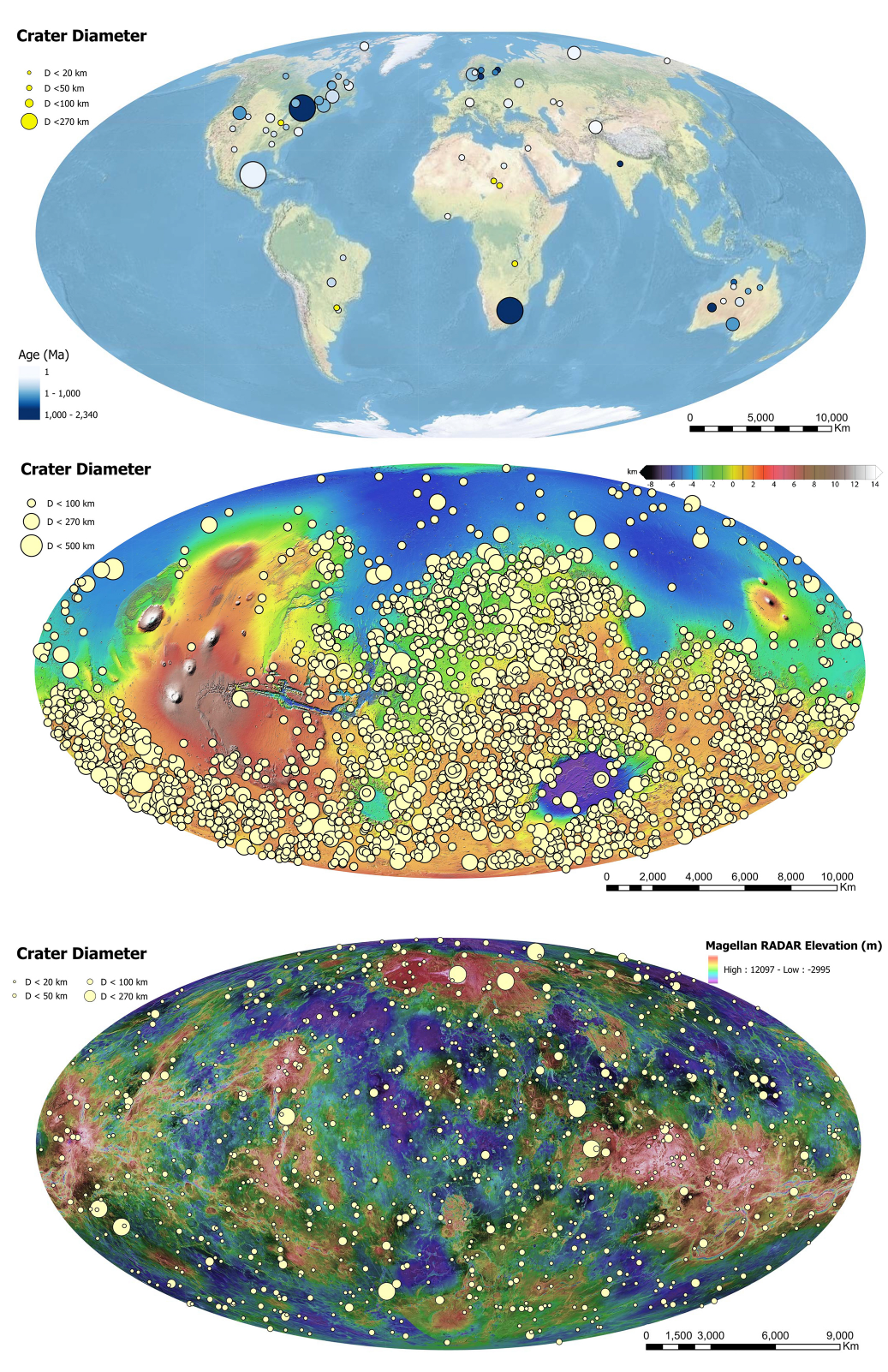}
\caption{(top) Map of Earth impact craters larger than 5 km diameter together with their respective ages coded by color. Filled yellow circles indicate an uncertain age. (middle) Map of Mars impact craters larger than 20 km diameter. Smaller impact craters are not shown for visibility. (bottom) Map of Venus impact craters larger than 5 km diameter. Venus impact craters are thought to be relatively young (less than 1 Ga) but relative and absolute ages are uncertain. 
Earth topography credit: Copyright:\copyright 2013 National Geographic Society, i-cubed, published by Esri for use in ArcGIS. Earth crater database adapted from \citep{mazrouei2019earth} and \citep{spray2018earth}.
Mars MOLA topography map: WMS server used for JMARS client by Arizona State University for Jmars \citep{christensen2009jmars}, credit: ASU, NASA, Goddard. Mars crater database from \citet{robbins2012new}.
Venus map Credit: Credit: USGS/NASA/JPL Magellan Mission, \url{http://planetarymaps.usgs.gov/cgi-bin/mapserv?map=/maps/venus/venus_simp_cyl.map}. Venus crater database is from \citep{herrick1997morphology}, and available at \url{https://www.lpi.usra.edu/resources/vc/}.}
\label{fig:impE}
\end{figure}

The interior of present-day Earth is convecting, and heat is lost efficiently by plate tectonics. Evidence suggests that modern-day plate tectonics has existed in its current form for at least 1 Gyr \citep{stern2005}, with suggested onsets at least 3-4 Gyr ago \citep[e.g.][]{dhuime2015,cawood2006,nutman2002}. The clear topographic dichotomy between oceanic and continental crust, with the majority of the volcanic activity localized at the plate boundaries, is specific to plate tectonics. Earth's tectonic regime is also characterized by the subduction of oceanic lithosphere that enables the large-scale recycling of surface material into the mantle. Earth's tectonic processes and erosion combined are also responsible for the small number of impact craters to be found on Earth (see Figure \ref{fig:impE}, top panel).

This convection regime has not been observed anywhere else on present-day terrestrial planets. Mars is currently in a stagnant-lid regime. On Mars, the lack of tectonic activity and limited erosion due the thin atmosphere result in a heavily cratered, mostly old surface (see Figure \ref{fig:impE}, middle panel). It should be noted that stagnant lid does not mean geologically dead, since Mars appears to have been operating in the stagnant-lid regime for billions of years and still shows recent signs of activity \citep[e.g.][]{hartmann2005,hauber2011}. Other planets, like Mercury, could also be in stagnant lid, but closer to being geologically dead bodies \citep[see][]{grott2011}.

The current convection regime active on Venus is uncertain: while Venus does not display global plate tectonics, the planet shows signs of deformation, volcanic and tectonic activity, and lateral variation \citep[e.g.][]{basilevsky2003,Rolf2022,herrick2023,hanmer2023}, which would be out of place in a fully stagnant-lid regime, such as observed on Mars. As a result, the term ``single plate planet'' rather than ``stagnant lid planet'' has been used to describe Venus. It has been suggested that Venus could be an episodic lid planet, with brief overturns of the lithosphere into the mantle, separated by long stagnant periods \citep{turcotte1993,armann2012}. The ``plutonic-squishy lid'' regime has also been proposed, and could possibly explain the relatively young surface age, deformation and heat flux measurements better than relying on a ``well timed'' overturn \citep{Lourenco-Rozel2023,Rolf2022}. However, no definitive evidence has been found yet to settle the question. Venus' stronger tectonic activity compared to Mars can be also seen in the relatively small number of impact craters on its surface (see Figure \ref{fig:impE}, bottom panel) as erosion processes are negligible on this planet. 

Evidence for Mars' or Venus' ancient convection regimes is even more difficult to identify. Mars's surface is old and indicates that its most active phase occurred during the first billion year of its evolution \citep{carr2007surface}, however the available evidence does not shed much light on what the planet's early dynamics looked like.
Venus' relatively young surface does not allow any direct insight into its distant past, as older surfaces have been covered by more recent basaltic material. Indirect evidence from the atmosphere of the planet indicate that large-scale outgassing may have been required to generate the current volatile inventory \citep{weller2022,gillmann2022}, which could indicate an active regime like a mobile lid in the past. Such an active phase may not be required when considering other sources of volatiles like impact delivery \citep{sakuraba2019impact,gillmann2020} or could contradict $^{40}$Ar measurements unless outgassing occurs very early \citep{kaula1999,gillmann2022}.

\subsection{Volcanic production}
While it is obvious that Earth is volcanically active at present-day, the same cannot be said about Mars and Venus. However, dating of lava flows on Mars indicates volcanic activity in specific areas in the last few hundred million years \citep{Neukum2004,Werner2009,Carr-Head2010,Niles2010} and marsquakes in the Cerberus Fossa region indicate the presence of magma in the martian underground \citep{stahler2022}. On Venus, transient IR anomalies localized close to identified lava flows and surface changes in volcanic regions have been interpreted as current volcanic activity \citep{Smrekar2010,Herrick-Hensley2023}. Global volcanic production rates are estimated to be around 20-30 km$^3$/yr for Earth and the extrusive volcanic production to be about 4 km$^3$/yr \citep{Crisp1984,morgan1998}. In the case of Venus, it has been estimated to range from 0.3 to 10 km$^3$/yr \citep{fegley1989,bullock2001} to $\leq$ 2 km$^3$/yr \citep{head1991}. Estimates for Mars indicate lower production rates around 0.3 km$^3$/yr \citep{greeley1991}.

Past volcanic production is even more difficult to constrain, but all three planets show signs of an active past. Mars boasts some of the largest volcanoes in the solar system and multiple volcanic provinces. Venus' surface is mainly covered by basaltic plains, its mostly young surface indicates active resurfacing, and extremely long lava flows have been observed \citep{fegley2014venus,conrad2023}.

It is likely that the three planets were more active during the first billion years as they lost accretion heat. Furthermore, heating by short-lived $^{26}$Al might have played a role in the early evolution of the growing planetary embryos \citep{Lichtenberg2019}. Mars is considered to have completed accretion and core formation already within $\sim$ 10 Myrs \citep{Dauphas-Pourmand2011,Mezger2013} and magma ocean crystallisation occurred within $\sim$ 20 Myrs after the start of the solar system \citep{Bouvier2018}. In comparison to Earth-like planets, smaller bodies such as Mars would have lost heat more rapidly during their early stages of evolution. This could explain differences in estimated volcanic production rates. However, once a stagnant lid is in place, the planet loses heat inefficiently, which may be why Mars' core appears to still be mostly liquid \citep{irving2023first}. It could also cause the mantle to heat up and cause widespread partial melting (which is in turn countered by increasing thick lids; \citep{o2014}. Differences between Venus and Earth may originate from other characteristics, since their sizes are similar.

\subsection{Outgassing as evidence for planetary history}

The volcanic history of a terrestrial planet is critical to the evolution of its atmosphere. Volcanic outgassing is one of the major sources of volatiles for the fluid envelopes of a rocky planet. In general, a higher production of volcanic material is expected to increase global volatile extraction from the mantle, but transport from the interior to the atmosphere is modulated by interior and surface conditions \citep[see][]{gillmann2022}. For example, the specific composition and mass of outgassed species depends on interior composition, redox state \citep{hirschmann2008,ortenzi2020} and $P$-$T$ conditions, both in the mantle (during melting) and at the surface (upon release, \citet{gaillard2014}). 

As a rule, due to their size and heat budget, Venus and Earth are expected to produce more melt than Mars and possibly outgas more. Differences between Earth and Venus would be caused by past history and mantle dynamics differences: at present Venus is expected to experience less volcanism, but precise estimates are difficult to make. Based on their atmosphere composition all three planet would release CO$_2$. Section \ref{sub:volatiles} discusses the consequences for the volatile inventory. Mars could outgas CO$_2$ at low volatile concentrations due to its composition and reduced interior \citep{hirschmann2008}. It is unknown what the composition of Venus' outgassing is, but it is thought that water may not be released into the atmosphere because the planet's mantle is dry and/or the surface pressure is high \citep{gaillard2014}. 

The concentration of radioactive Ar in the atmosphere has been suggested to be a proxy for the volcanic history of a terrestrial planet. As potassium decays into $^{40}$Ar, argon partitions into the liquid phase during partial melting and is outgassed into the atmosphere. Based on assumed initial compositions for the planet (i.e. K/U ratio), one can compare the relative outgassing of the planets. Measurements for $^{40}$Ar suggest that Earth could be $\approx$ 50\% degassed, and Venus would have outgassed $\approx$ 25\% of its volatiles, making its atmosphere more primordial \citep[less affected by outgassed volatiles, at least after the end of the magma ocean phase,][]{kaula1999}. The uncertainties in initial compositions however, make this number poorly constrained \citep{gillmann2022}. Additionally, the timing of the outgassing is important, as an early massive outgassing, when little $^{40}$Ar has been produced, could possibly release large amounts of volatiles without being mirrored by $^{40}$Ar measurements. Finally, especially in the case of Mars, losses of Argon through atmospheric escape could be difficult to quantify \citep[i.e.][]{hutchins1996,gillmann2011,leblanc2012}. However, models that track the evolution of outgassing and $^{40}$Ar production under a continuous stagnant lid regime throughout the simulated history of Mars \citep{ONeill2007,gillmann2011} produce volcanic fluxes that are consistent with present-day $^{40}$Ar measurements \citep{pollack1982}. In the end, though qualitative, differences in the records of Earth and Venus point at a different style of outgassing, which could be explained if Earth operated mostly under mobile lid/plate tectonics (associated with larger outgassing), while Venus has had an episodic regime \citep{o2014}, a change in its regime \citep{Gillmann-Tackley2014}, or if its mobile-lid/plate-tectonics regime is limited to its early evolution \citep{arkani1993,Gillmann-Tackley2014,weller2022}.

The generation (or not) of a magnetic field in the planetary core is also a clue pointing towards different evolutionary paths. Earth shows a self-sustained magnetic field with a strength similar to present-day since at least 3.2 Ga, possibly 3.45 Ga \citep{tarduno2014detecting}. Mars, however, sustains no dynamo at present-day but traces of ancient remanent magnetisation of its crust exist and date back to the first 600 Myr of its evolution \citep{Steele2023}. Finally, Venus shows no sign of a global self-sustained magnetic field at present-day \citep{Phillips-Russell1987} and no remanent magnetisation has been observed so far.
A magnetic dynamo is generated by convection in the liquid layer of the metallic core. It can be thermal (sufficient heat is lost at the core mantle boundary to drive core convection) and/or chemical (iron solidifies and sinks towards the solid inner core, thus stirring the liquid core), but it is always related to heat extraction by mantle convection and core cooling. The lack of a dynamo is thus indicative of either a completely frozen core, a stably stratified core \citep{Jacobson2017} or a too low heat flux out of the core (insufficient to cause core convection, thus no thermal dynamo). 

The lack of a dynamo on Venus is especially puzzling if one assumes that Earth and Venus share a similar core composition and characteristics, as simulations would predict that slow rotating present-day Venus could generate a detectable magnetic field \citep[i.e.][]{ORourke2018}. The current estimates for the tidal Love number on Venus are not precise enough to discriminate between a solid and liquid core \citep{dumoulin2017}. It has been suggested that Venus did not experience giant collisions such as Earth's moon-forming impact, and therefore retained a stratified core \citep{Jacobson2017}, thus explaining the lack of magnetic field. In that case, the slow retrograde rotation observed for Venus would be caused by tidal interaction \citep{dobrovolskis1980,Correia-Laskar2003,leconte2015}, not impact despinning. However, it has also been proposed that impacts were the cause of the peculiar Venus rotation \citep[i.e.][]{mccord1968,Alemi-Stevenson2006,Raymond2013}. If the core is unstratified, the lack of a present-day magnetic field would suggest near-complete solidification or that the thermal conductivity of Venus' core material is higher than generally assumed \citep{ORourke2018}.

\subsection{The role of impacts in interior evolution}
Impacts have also been suggested to affect the interior dynamics of terrestrial planets. During a collision, the kinetic energy of the impactor can be transferred to the target planet's mantle as thermal energy. If the impact is energetic enough and large enough (so that the heated volume extends beyond the lithosphere and into the mantle, typically for impactors above 100 km radius), then the mantle dynamics can be affected by the collision \citep{gillmann2016}. 

Such large impacts have been shown to cause temporary subduction events in numerical simulations of Venus and Earth \citep{gillmann2016,ONeill2017,Borgeat-Tackley2022} and to increase lithosphere mobility. However, their consequences in simulations are relatively short-term (a few hundred million years at most) and any mobile lid regime they cause would have naturally emerged within the simulation parameters \citep{Borgeat-Tackley2022}. 

Simulations of Mars' largest recorded impacts have also failed to result in any lasting change for both climate or internal state \citep{turbet2020,ruedas2019}. Only early-on ($\ge$ 4 Ga) impactors are likely to be large enough (above a few hundred kilometer radius, in order to significantly penetrate into the mantle) to have affected the interior and climate significantly on the long-term. Finally, it is unclear how different the impact flux would have been on the three terrestrial planets and what marker would allow us to distinguish between possible scenarios \citep[e.g.][]{gillmann2020}.

\subsection{Modelling efforts and governing parameters}
While numerical models are now capable of simulating Earth-like single sided plate tectonics, where only one of the plates sinks back into the mantle even when both are oceanic plates \citep[i.e.][]{Crameri-Tackley2014,Crameri-Tackley2015}, the actual conditions for a planet to develop this regime are more difficult to assess. In essence, it requires the lithosphere to be brittle enough so it can break rather than flow in a ductile way, and not so strong that convective stress can never break it. In terms of modelling, this means that the yield strength should be low \citep[a few tens of MPa - see for example][]{Lourenco-Rozel2023}. However, these values do not fit with laboratory measurements for Earth materials, which are reportedly much higher, potentially because these experiments also do not account for localized weak zones, faults and lateral material variations occurring on larger length scales. 

Temperature is a major parameter that affects rheology. It has been shown that surface temperature variations, due to the evolution of the climate, could have direct consequences on the convection regime. For example, \citet{Lenardic2008} have shown that a surface temperature increase of about 100 K could shut down existing plate tectonics in an Earth-like model by reducing the viscosity of the convective mantle below the lithosphere. Because of the exponential dependence of the viscosity upon temperature, increasing temperature by 100 K reduces the viscosity by about one order of magnitude \citep{kohlstedt1995}. In turn, lower viscosities would reduce convective stresses upon the lithosphere \citep{moresi1998}. If convective stress drops below the yield strength of the lithosphere, plate tectonics stops \citep{moresi1998}. Models of Venus \citep{Gillmann-Tackley2014} indicate that water abundance variations in the atmosphere of Venus could change surface temperatures by several 10s to 100s K and cause a switch from a mobile to a episodic lid regime as partial pressure of water evolves towards its present-day value. \cite{Noack2012} highlight that particularly high surface temperatures (higher than at present-day) could reduce lithosphere viscosity enough to allow it to be dragged by the mantle and become mobile. Such high temperatures (and possibly present-day Venus temperatures) could also lead to a relatively ductile lithosphere and prevent a plate-like behaviour by allowing faults to ``heal'' more easily \citep{bercovici2014}. It is still unclear whether any of those scenarios occurred during the evolution of Venus.

The temperature within the lithosphere can also be affected by intrusive volcanism, in turn possibly favored by high surface temperatures \citep{Lourenco2020}. The phenomenon can be local and generate weak zones or be more widespread and lead to a warm and soft lithosphere.

Water has also been suggested to play an important role in the rheology of the mantle and lithosphere of terrestrial planet due to its effect on the friction mechanisms and pore pressure \citep{korenaga2013}. Through the dependence of viscosity on water, the water cycle affects mantle dynamics. The correlation between plate tectonics and liquid surface water in the solar system has motivated this line of thought. If water is needed for plate tectonics \citep[i.e.][]{Solomatov2004}, a dry early state, as described before, would favor the single plate/episodic lid regime as suggested for present-day Venus. On the other hand, low water abundances may lead to a higher mantle viscosity \citep{xie2004}, which could, for example, improve the coupling between the mantle and lithosphere. Because the abundance of water in the mantle is also dependent on mantle dynamics, a feedback mechanism would exist between thermal evolution and the water cycle \citep{crowley2011relative}.

In the end, despite important advances in our understanding of different regimes and mechanisms at work in planetary mantles, the lack of actual knowledge regarding the state and conditions inside solar system terrestrial planets prevents us from proposing a definitive explanation for the differences we observe. This is in part because we have little direct knowledge of the rheology and water content of both Mars and Venus. In recent years, partly in preparation for future exoplanet atmosphere characterization, efforts have been made to use atmosphere composition to investigate interior evolution, by using the consequences of outgassing integrated over the planets' history e.g. \citep{gillmann2020,krissansen2021,ortenzi2020,weller2022,warren2023}.

\subsection{Consequences on volatile inventory}
\label{sub:volatiles}
The convection regime of all three terrestrial planets had important consequences for their whole evolution and surface conditions. On Earth, the composition of the atmosphere is the result of many processes, collectively referred to as volatile cycles \citep[see][]{dasgupta2010deep}. Over geological timescales, surface solid-gas reactions, the effect of life and the recycling flux at the subduction zones governs, for example, the abundance of CO$_2$ in the atmosphere, with complex and uncertain feedbacks \citep[e.g.][]{friedlingstein2006}, while on short timescales, ocean-atmosphere-biosphere interaction would dominate. 

Comparatively, volatile exchanges are likely to be much simpler on Mars and Venus \citep[e.g.][]{van2007water}. The long-term stagnant lid on Mars makes any recycling of gaseous species into the mantle unlikely. Same applies for Venus, at least since the time of the formation of the present-day surface and atmosphere. Additionally, at present-day Venus surface conditions, carbonates are not stable and, due to the lack of surface water, can not form \citep{fegley2014venus}. A limited form of volatile storage can act as a temporary sink even on stagnant lid planets. However, with no volatile cycle and no pathway back into the mantle, it can be easily destabilized. \cite{Honing2021} have described such an evolution and suggest that volatiles would be released back into the atmosphere during the first billion year as temperature increases either at the surface or at the depth where the material is buried.

In the absence of a stabilizing feedback, volatiles would accumulate in the atmosphere when not lost by escape processes, which would be the situation we observe for Mars and Venus relative to Earth \citep{bean2017,turbet2019}. The nature and amount of gases produced by partial melting and volcanism, then released into the atmosphere, are primarily dependent on the composition and redox state of the mantle, which both remain uncertain, even for Earth. The redox state of the mantle can be expressed using the oxygen fugacity measured as the ratio between the abundances of the different iron ions relative to a buffer, like the IW buffer (see also Section \ref{sec:MO}). Earth's upper mantle is, despite some heterogeneities, rather oxidized at present-day, at approximately IW+1 to IW+5 \citep[e.g.][]{ballhaus1990,Hirschmann22GCA}, while Mars is more reduced, around IW+0 to IW+1 \citep{herd2002,shearer2006,karner2007,hirschmann2008}, and Venus remains unconstrained, possibly close to Earth \citep{schaefer2017}. As a result, outgassing on Mars may be characterized by lower CO$_2$ concentrations in the magma than on Earth and, possibly, Venus \citep{hirschmann2008}. 

However, it has also been suggested that surface conditions, such as surface pressure, may affect the composition of the outgassing into the atmosphere. For example, surface pressure affects the gas-gas and gas-melt equilibria, which modifies volatile speciation during release \citep[i.e.][]{gaillard2014}. As a result, high surface pressures (above $\approx$10 bar) may inhibit part of water outgassing into the atmosphere, increasing its solubility in the lava. \citet{gaillard2014} suggest $\approx$90$\%$ of the initial water content of lava could be trapped. Instead, more Earth-like surface pressures would lead to water being the dominant outgassed species. On the other hand, CO$_2$ can outgas even at pressures of $\approx$100 bar. Venus-like surface pressures would prevent most H$_2$O from reaching the atmosphere under present-day Venus conditions, and possibly in the past, as long as surface pressures remain high (above a few bar). Earth may thus occupy a niche where the mantle is oxidized enough and the surface pressure low enough to allow for the outgassing of both CO$_2$ and water. Under Mars- or Venus-like conditions, either the mantle or at the surface, respectively, could stifle the release of one or both of those species \citep{gillmann2022}.

Episodic volcanic outgassing is also often suggested for Venus. Based on the most recent resurfacing event that covered most of the surface with basalt (or the equivalent production under the equilibrium hypothesis; \citet{phillips1992}; see also \citet{smrekar2018venus}, \citet{lopez1998progressive} estimate that 9-10 such events would be required to build up the present-day CO$_2$ atmosphere, assuming an Earth-like lava composition. This corresponds to a total amount of extrusive lava of about 5$\times$10$^{10}$ km$^3$ (based on possibly generous estimates from \citep{bullock2001}), which would translate into a total volume of melt of 2.5-5$\times$10$^{11}$ km$^3$. As a reference, the volume of Venus' mantle is about 7.8$\times$10$^{11}$ km$^3$. Therefore, episodic production of the atmosphere requires massive outgassing, which could be in conflict with $^{40}$Ar measurements, unless it occurred early or a large fraction of the atmosphere was already acquired during accretion or from other sources (impacts, destabilization of surface material).

Based on a detailed survey of the surface, \citet{head2022detecting} suggest that 80-100 (likely smaller scale) resurfacing events would be required and that the atmosphere of Venus is therefore ancient, which seems to be supported by $^{40}$Ar measurements. Simulations that reproduce episodic evolution \citep{armann2012,Gillmann-Tackley2014,gillmann2020,weller2022} require multiple billions of years to generate enough volcanic episodes to produce Venus' 90 bar CO$_2$ atmosphere, or fail to generate enough large events, due to the large amount of heat required for such large melting episodes - heat that is extracted from the mantle during the process. 

Additionally, if CO$_2$ would outgas easily (provided the mantle is oxidized enough), water may have been unable to reach the atmosphere for a large part of the Venus evolution due to high surface pressures \citep{gaillard2014, Gaillard2022}.
\cite{way2022large} propose that such large-scale volcanism (possibly akin to Earth's Large Igneous Provinces) may have contributed to the present harsh venusian climate. However, little is known about the ancient volcanic production and outgassing on Venus, and modelling based on existing constraints, like $^{40}$Ar, noble gases, D/H ratio is an active and promising avenue of research \citep{Noack2012,gillmann2020,krissansen2021}.

\section{Atmospheric loss}\label{sec:atm-loss}

Looking at the different atmosphere masses on Earth, Mars and Venus, signs of volatile losses can for instance be inferred from the fractionation of non-radiogenic gases and the deuterium to hydrogen ratio (D/H). Indeed, Mars has been observed to be more enriched in deuterium by a factor of 5 than Earth \citep{owen1988deuterium} and Venus by a factor of 150 \citep{donahue1997ion}. The preferential loss of the lighter isotope compared to the heavier one affects the ratio between the two isotopes - the more a sample deviates from the primordial ratio, the stronger the loss. One should however remember that this is not true for all loss mechanisms (impact erosion of the atmosphere or intense hydrodynamic escape would not fractionate H, for example), and that quantifying that loss \citep{Pahlevan+19EPSL,kurokawa2018lower,zahnle2023} or how the ratio is affected by other processes is still uncertain \citep{grinspoon1993implications,avice_noble_2022}. 

Generally speaking, three main categories of escape are considered and discussed below: hydrodynamic escape (or thermal escape, as it is due to heating of the atmosphere), non-thermal escape (chemical reactions and collisions with particles from the solar wind) and impact erosion (caused by meteoritic collisions).

Atmospheric escape fluxes have been directly measured or calculated for the three planets. Additionally, Mars and Venus appear drier than the Earth, and water plays an important role in the long-term terrestrial planet evolution. Therefore, it is no surprise that atmospheric loss processes have been proposed as a possible cause of the diverging evolutionary pathways for terrestrial planets in the Solar System. 


Noble gases in the Venus atmosphere have been analyzed by mass spectrometers and gas chromatographs on board the Pioneer Venus and Venera probes. Although a similar pattern is observed for the three terrestrial planets, the $^{20}$Ne/$^{36}$Ar ratio is depleted by a factor of $\approx$ 3 in the Venus atmosphere relative to Earth's. Ne and Ar are more abundant in Venus' atmosphere than in Earth's atmosphere, by at least factors of 10 and 50, respectively. $^{20}$Ne and $^{36}$Ar abundances are close to chondritic for Venus, but depleted for Mars and Earth, as are $^{84}$Kr and $^{130}$Xe for those two planets. In addition, $^{130}$Xe/$^{84}$Kr is measured at one order of magnitude below the chondritic ratio, indicating further loss \citep{pepin1991origin}. In the case of Venus, however, Kr and Xe are still poorly constrained, thus so far it is not possible to assess whether or not Venus follows the Earth-like pattern \citep{lammer2018}. With the caveat of Venus Xe measurements, all three planets are depleted in noble gases compared to the solar composition. In general though, Venus' noble gas elemental spectrum is suggested to be more solar-like than those of Earth and Mars.

On Mars, the isotopic components of atmospheric Kr are close to those of the solar wind and remained mainly unfractionated \citep[e.g.][]{pepin1991origin,becker2003isotopic,conrad2016situ}. Xe isotopes in Earth’s and Mars' atmosphere are enriched in heavier isotopes. The case of Xe is still actively debated, with the fractionation possibly attributed to long-term (4.5-2 Ga; \citet{avice2020}) escape of ionized hydrogen dragging of Xe \citep{zahnle2019strange}. 

Neon is less fractionated \citep{wieler2002noble} in Venus atmosphere ($^{20}$Ne/$^{22}$N = 11.9 $\pm$ 0.7) than in Earth's atmosphere ($^{20}$Ne/$^{22}$Ne = 9.8) with respect to solar composition ($^{20}$Ne/$^{22}$Ne = 13.5). Because Ne is only slightly heavier than O, hydrodynamic escape of Ne implies hydrodynamic escape of oxygen, and Ne may be used as a tracer of oxygen escape at primitive stages \citep[e.g.][]{gillmann2009b}. There is also no evidence of changes in Ne, Ar and Kr isotopic compositions (non-radiogenic) during the past 3.5 Ga on Earth \citep{avice2017origin}. On all three planets it has generally been suggested that their lighter noble gases were fractionated during an early hydrodynamic escape phase (the first few hundred million years).

\subsection{Hydrodynamic escape}
Hydrodynamic (or thermal) escape occurs due to heating of the atmosphere by intense EUV radiation from the Sun or by radiation from a primitive hot rocky planet \citep["core-powered"]{ginzburg2018}. It causes atoms in the upper atmosphere to exceed the escape velocity, leading to very efficient escape of the bulk gas of the upper atmosphere where the atmosphere remains a collisional fluid as it passes through the transonic point \citep{chamberlain1963planetary,chassefiere1996hydrodynamic,fossati2017aeronomical}. Light species (especially H) are also the most efficient at escaping via this process, as they require less energy to reach bulk escape velocity. 

Heavier species (such as helium, carbon or oxygen atoms) can undergo escape via so-called hydrodynamic drag where they are carried along with the stream of lighter hydrogen atoms \citep[e.g.][]{hunten1973escape}. Due to solar evolution (the decrease in EUV flux with time \citep{claire2012}, this implies the most favourable conditions for such a process occur during the first few hundred million years. At lower energy levels, a transition occurs from hydrodynamic escape to the much less efficient Jeans escape, causing the loss of atoms populating the high energy tail of a Maxwell distribution \citep{jeans1925dynamical,chamberlain1963planetary}, and still occurring at present-day on Earth, Mars and Venus.

Hydrodynamic escape would have efficiently removed the primordial H/He atmosphere that all three planets obtained from the solar nebula during their formation \citep{lammer2018,lammer2020}. Earth's D/H seems to be incompatible with oceans originating from the solar nebula \citep{marty2012}. Then, during the magma ocean phase, the ongoing thermal loss would have affected the water vapor/CO$_2$ atmosphere of Earth, Venus and Mars to various degrees, whether it came from interaction of the atmosphere with the magma ocean or from catastrophic outgassing during magma ocean solidification. Finally, if a planet, such as Venus, entered a runaway greenhouse state, then hydrodynamic escape would also have been a dominant factor for water loss. The loss of water would have remained efficient but the presence of more abundant heavier species would have led to diminished loss rates. Still, modelling efforts have shown that a Mars-like body orbiting at Earth-like distance from the Sun would lose its atmosphere (85 bar H$_2$O and 11 bar CO$_2$) in about 15 Myr \citep{Odert2018}, and that Venus would only take 50 Myr to lose about 2 Earth ocean equivalents \citep{gillmann2009}. Gravity, the composition of the atmosphere and the proximity to the star all affect the loss rates, with smaller or closer planets being more affected. Thus, Mars could have lost a large part of its initial water inventory and even CO$_2$ \citep{Tian+2009,Odert2018}. Mars being drier than Earth due to early escape ($\ge$ 4 Ga) is consistent with martian meteorites measurements \citep{cassata2017meteorite}.

The planets are efficiently dried out, as long as a supply of water is available in the atmosphere. \cite{Kasting&Pollack83Icarus} calculated that hydrodynamic escape was the dominant process as long as the water mass mixing ratio of water in the lower atmosphere exceeded 0.1, even during the later evolution. 

In short, factors that favor a strong hydrodynamic escape are (in order of importance): (i) an H-rich atmosphere (water vapour for instance), (ii) a strong EUV flux (for example during the first few hundred million years of the evolution) and (iii) a small planet with low gravity \citep[see also][]{catling2009}. Because of (i) and (ii) thermal escape is likely the dominant process during the first few hundred million years of planetary evolution. The escape of lighter species (H) is favored, but a strong enough flux can cause the loss of heavier components. Conversely a CO$_2$ atmosphere, low EUV flux and stronger gravity would reduce escape rates, while magnetic fields do not impede hydrodynamic escape.

\subsection{Non-thermal escape}
Non-thermal escape covers a few mechanisms that involve the interaction of the upper atmosphere molecules (above the exobase) and ions with energy and particles brought by the solar flux, such as: charge exchange, photochemical dissociation, sputtering, polar wind, plasma clouds. In each case, the interaction leads specific molecules to reach high enough velocities to escape \citep{catling2009}.

Over the last decades, a large number of studies have attempted to quantify the escape rates of heavy ions (such as O$^+$, O$_2^+$ and CO$_2^+$) on Mars, Earth and Venus based on observations. \citep{ramstad2021} have summarized these results (their figure 5) and found that estimates for Earth, Mars and Venus tend to be roughly similar, despite the large variations in planetary size, atmosphere composition, presence (or absence) of a global magnetic field. Mars' heavy ion escape rates range between 2 $\times$ 10$^{23}$ s$^{-1}$ and 10$^{25}$ s$^{-1}$; for Venus reported rates cover 1-6 $\times$ 10$^{24}$ s$^{-1}$ (with an outlier one order of magnitude above), while Earth's escape fluxes are higher: 7 $\times$ 10$^{24}$ to 10$^{26}$ s$^{-1}$. For O$^+$ alone, rates ranges from 2-4 $\times$ 10$^{24}$ s$^{-1}$ for Mars, to 1-4 $\times$ 10$^{24}$ s$^{-1}$ for Venus and 5 $\times$ 10$^{24}$ s$^{-1}$ to 5 $\times$ 10$^{25}$ s$^{-1}$ for Earth \citep[see][and references therin]{ramstad2021}. 

These reported values also cast doubt on the previously accepted notion that magnetic fields provide an efficient shield against atmospheric escape, which in turn was used to explain why Earth was so different compared to Mars and Venus, and retained its water. To explain the similar measured escape rates reported above, it has been suggested that Earth's magnetic field extends much further into space than its atmosphere (a few Earth radii) and therefore intercepts more energy from the Sun than both Mars or Venus \citep{brain2012planetary}, leading to increased atmospheric escape. 

It has also been suggested that magnetic fields could actually enhance atmospheric escape \citep[e.g.][]{gunell2018intrinsic}. This idea has been disputed by \citet{tarduno2014detecting}, for example for underestimating the return fluxes (ionized molecules escaping following magnetic lines that reconnect to the planet rather than open lines in the magnetotail). The current state of the research indicates that the magnetic field affects escape mechanisms (a magnetized planet will lose molecules at the cusps of the field rather than in the tail of the induced field) and affect which mechanism can be dominant. However, so far we are not able to explain exactly what the resulting escape flux will be \citep{gronoff2020atmospheric}. As a consequence, it remains uncertain by how much the presence of a past magnetic field could have affected the respective evolution of Mars, Venus and Earth, although as discussed above there may be hints that the presence of a magnetic field does not result in a radically different volatile loss history.

\subsection{Past escape reconstruction}
Attempts have been made to reconstruct past escape rates \citep{hunten1989escape} based on the present-day measurements for Mars \citep{chassefiere2004mars,gillmann2009}, with the latest contributions being the MAVEN results \citep{jakosky2021atmospheric}, and Venus \citep{Gillmann-Tackley2014,Persson+20JGR}. These are obtained by extrapolating the present-day escape rates obtained at high and low solar EUV fluxes during a solar cycle. However, they ignore changes in atmosphere composition and structure, or exosphere temperature and remain dependent on other factors, such as the existence of a magnetic field \citep{gronoff2020atmospheric,way2022searching}. Depending on the specific conditions at a given time, escape rates caused by a given mechanism are likely to vary and the dominant escape process would certainly change: water vapour atmospheres would be defined by strong H loss from hydrodynamic escape, while dry atmospheres would see non-thermal processes gain importance. Generally speaking, the two types of atmospheres are often used to describe, respectively, (1) the primitive phase of planetary evolution during magma evolution or runaway greenhouse, and (2) the last $\approx$ 4 Gyr and present-day.

For example, CO$_2$ in the thermosphere radiates energy back to space, leading to infrared cooling of the upper atmosphere and affecting the structure of the atmosphere \citep{roble1989will}. This is why Venus' exobase is both cooler and occurs at a lower altitude than Earth's. A hot, expanded atmosphere, resulting from a switch from a CO$_2$-rich atmosphere to a N$_2$ atmosphere for example, could possibly lead to escape rates orders of magnitude higher \citep[even though estimates are very uncertain;]{lammer2006,way2022searching}. Past reconstructed escape rates should thus be considered with caution, only as a first order approximation. They are currently, however, our best shot at ancient loss rates, since we are not yet able to predict a total escape rate from a given set of conditions. 

Non-thermal escape is modulated by (i) a planet's gravity, with weaker gravity leaving planets more vulnerable to loss. Non-thermal escape is also a function of (ii) the composition of the atmosphere, since most mechanisms depend on chemistry. However the precise effects of different atmosphere composition on loss rates is still uncertain. Finally, as a large part of the escape involves ions, (iii) magnetic fields existence and configuration will affect non-thermal escape. Non-thermal escape is dominant during late evolution and for species such as O (all planets) or C, N, Ar (for Mars). 

\subsection{Loss of water and the fate of remnant oxygen}

From observation, for non-thermal processes, the ratio of the escape of H and O appears to be close to or below the stoichiometry of water \citep[between 1.1 and 2.6][for instance, in the case of Venus]{persson2018}. Such a ratio leads to the full escape of water or a moderate loss of O relative to the source water. On the other hand, thermal escape is expected to lead to a relative accumulation of oxygen in the atmosphere due to the strong differences in loss efficiencies for species of different masses. Oxygen, being heavier, is lost at a much reduced rate compared to H. The ratio of the loss of H to O would thus be much higher than that of their respective abundance in the atmosphere \citep[2 H for 1 O, assuming they come from the photodissociation of water;][]{Kasting&Pollack83Icarus,gillmann2009b}. As a result the atmosphere becomes more oxidized unless the excess O can be removed through other processes such as oxidizing reducing gas species and atmosphere-surface reactions. The fate of the remnant O has been considered to be a constraint for the evolution of the Venusian atmosphere, and is discussed in detail in Section \ref{sec:venus}. On Earth, reconstructed atmosphere compositions dating back from before the Great Oxygenation Event \citep{lyons2014,catling2020} indicate that a long-lived O$_2$-rich atmosphere resulting from the process mentioned above is unlikely. Banded iron formations \citep[][and references therein]{Canfield2005}, for example are on Earth indicators of oxidation of iron in oceans under an atmosphere with low O$_2$ concentrations and low seawater sulfates \citep[and provide a link to the consequences of life for planetary evolution, see below, ][]{cloud1968atmospheric}.

As mentioned above (see Section \ref{sec:atm-loss}), signs for long-term water loss are evidenced by the D/H measurements on Earth, Mars and Venus. \citet{alsaeed2019}, for example, reconstruct the evolution of water reservoirs based on atmospheric escape on Mars and suggest that at 3.3 Ga, a global equivalent layer of 40-120 m of water could have existed to explain the present-day D/H ratio. Further, Mars' atmosphere is enriched in $^{15}$N over $^{14}$N, which has been suggested to result from non-thermal loss processes \citep{shizgal1996nonthermal,gillmann2011}. On the other hand, non-thermal escape on Mars is unable to account for the loss of a water equivalent above a few 100 mbar over 3-4 Gyr \citep{gillmann2009,gillmann2011}. At the same time, the cumulated water escape on Venus over the last 3.8 Gyr was recently estimated to amount to a few centimeters to tens of centimeters only \citep{Persson+20JGR}.

It seems unlikely that non-thermal escape could, by itself, cause the divergences observed in the solar system. However, it can critically influence already thin atmospheres of low mass terrestrial planets like Mars'. Additionally it remains a significant evolutionary process that shaped the observed characteristics of thicker atmospheres like Earth's and Venus'. 

\subsection{Atmosphere erosion by impacts}
Impact-related erosion of the atmosphere can lead to volatile loss\citep{Cameron1983,Melosh-Vickery1989}. Four main mechanisms can cause escape \citep[e.g.][]{dehant2019geoscience}: (i) direct ejection of the atmosphere by the shock wave, (ii) the rising of the vaporized projectile and sediments in a vapor plume, (iii) heating of the atmosphere by ejected particles and (iv) in the case of very large collisions, ground motion of the planet can accelerate particles of the atmosphere above escape velocities.
That last process is deemed to be the most efficient one. \cite{Genda-Abe2003} estimate that such an impact ($R_{\rm i} \sim$ 1000 km) could remove about 20\% of a preexisting atmosphere and the presence of a water ocean could even enhance the atmosphere loss \citep{Genda-Abe2005}. More recent work \citep{Schlichting2015}, derives estimates depending on relative mass of the impactor to the planet and its velocity. Significant loss (above 10\% of a preexisting atmosphere) requires the impactor to be traveling at several times the planet's escape velocity and/or to be of a comparable mass to the target, which is more easily achieved for Mars than for larger planets.

A key feature of impact erosion is that it generates bulk ejection of the atmosphere and does not isotopically fractionate individual elements in the planetary atmosphere, unlike other processes described above. It could, however fractionate the volatile budget of two elements, in case those elements are not distributed equally within the planet, for example if one resides preferentially within the atmosphere while the other one is trapped in the interior\citep[e.g.][]{Massol+16SSRv}. Impact erosion of the atmosphere is probably the main mean of removing Kr from the atmospheres of terrestrial planets, since Kr (generally non-fractionated) is too heavy and difficult to ionize to be significantly lost through the previously discussed thermal and non-thermal processes \citep{pepin1994,catling2009,cassata2022}.

Smaller impactors ($R_{\rm i}$ $\le$ 1000 km) individually remove less atmospheric volatiles through processes (i)-(iii) than larger ones, however, they are more efficient at atmospheric erosion per unit of mass. \citet{schlichting2018atmosphere} highlight that the much more numerous population of smaller impactors (1 km $\leq$ $R_{\rm i}$ $\leq$ 20 km), as a whole, can remove more volatiles than the few larger collisions (25 km $\leq$ $R_{\rm i}$ $\leq$ 1000 km). They argue that such a flux of small impactors would be able to erode an Earth-like atmosphere within a few billion years. While this can affect decisively thin atmospheres (such as Mars'), this escape rate makes little difference, by itself, on a planet with a multi-bar atmosphere such as present-day Venus. 

In short, volatile loss is very dependent on the planetary mass and impact velocity: less massive planets such as Mars are more vulnerable to erosion than Earth or Venus \citep{pham2011effects} and faster impactors can drive a greater loss. Furthermore, massive atmospheres would be only marginally affected by erosion, relative to their total masses, in contrast to planets with thinner atmospheres. Finally, impacts can also act as sources of volatiles depending on their exact composition \citep[][for example]{gillmann2020} and affect atmosphere chemistry \citep{schaefer2017}. While erosion, from a relatively predictable population of impactors, averages out, volatile deposition by collisions is much more susceptible to stochastic events (such as a single large volatile-rich impact), which may cause drastic changes in atmosphere bulk composition and chemistry.

\section{Unique features of the history of Mars, Venus and Earth}
\label{sec:feat}

\subsection{Mars} \label{sec:mars}

As Mars is less massive than Earth, magma ocean pressures on Mars would have been lower, compared to the Earth's magma ocean pressures. Geochemical studies on martian meteorites support core formation pressure of about 14~GPa to explain the depletion in siderophile and chalcophile elements in the martian mantle \citep[e.g.][]{Righter&Chabot2011}. Therefore, the iron disproportionation would have had a limited effect on Fe$^{3+}$ abundance and a $f$O$_2$ gradient in the martian magma ocean. The estimated surface $f$O$_2$ above the magma ocean is about IW-1$\sim$IW+1 \citep{Deng+20NatCom}. This $f$O$_2$ is high enough that a large fraction of H would be dissolved in the magma ocean as H$_2$O \citep{Gaillard2022}. As the orbit of Mars is far enough to escape from the runaway greenhouse state, Mars is expected to have crystallized rapidly, though the predicted duration of magma ocean crystallization shows a large discrepancy between thermal models \citep[$<\sim$1~Myr,][]{Elkins-Tanton08EPSL,Lebrun+13JGR} and geochemical constraints \citep[several to tens Myr after core formation,][]{Borg+2016,Krujier+17,Kruijer+2020}. Several processes such as late differentiation caused by cumulate overturn \citep[e.g.][]{Elkins-Tanton+2005} and heating by late accretion\cite[e.g.][]{Marchi+2018} have been proposed as potential causes of the prolonged Martian magma ocean. Depending on its outgassing efficiency, its water would remain at the surface or in its interior \citep{Odert2018}.

Mars' ancient surface shows a dichotomy between the southern highlands and the northern lowlands. The dichotomy is considered to be the oldest geological feature on Mars and was created within the first 400-500 Ma after the planet's formation \citep{Nimmo-Tanaka2005,Carr-Head2010}. Different processes have been suggested as the origin of this topographic feature like (i) impact excavation of the northern hemisphere \citep{Wilhelms-Squyres1984,Frey-Schultz1988,Andrews-Hanna2008,Marinova2008,Nimmo2008}, (ii) low degree convection \citep{Zhong-Zuber2001,Roberts-Zhong2006,Zhong2009,Keller-Tackley2009,Sramek-Zhong2010} and (iii) an impact in the southern hemisphere forming a so-called impact megadome \citep{Reese-Solomatov2006,Reese-Solomatov2010,Reese2010,Golabek2011,Ballantyne2023,Cheng2024}. Also it has been suggested that Mars could have developed short-term plate tectonics during its first few hundred million years \citep{Frey2006,Sautter2015}. 

In short, due to its relatively small mass, and possibly limited water inventory, Mars would have transitioned from a possibly active first billion year into a stagnant lid regime. At about the same time, it lost a large part of its initial water inventory and its magnetic field. Details regarding the climate on ancient Mars beyond increasing desiccation remain debated, and both "warm and wet" and "cold and wet/snowy" origins have been proposed \citep{wordsworth2015comparison}, as well as episodic warming by volcanism and reducing greenhouse gases \citep{halevy2014episodic}.

Observations of the martian surface give an idea of the consequences that hydrodynamic escape could have as Mars' terrain age allows studying its distant past. Fluvial features indicate a decline in water's role in shaping the surface of the planet, evolving from valley networks in the Noachian to episodic outflow channels in the Hesperian towards the more arid planet since the Amazonian \citep{carr2007surface}. Based on mineral alteration in the presence of diminishing amounts of water the subsurface mineralogy analysis suggests a similar evolution from Phyllosian to Theikian to Siderikian \citep{bibring2006}. 

Long-term evolution models suggest outgassing would be relatively limited, possibly due to interior redox conditions inherited from its primitive history \citep{hirschmann2008,Deng+20NatCom}. A large portion of the outgassing would have occurred during the first billion years of its evolution \citep{lammer2018}, before most of its interior heat was lost. However, even such low volatile release rates could be sufficient to affect surface conditions and the composition of the atmosphere. Based on the estimated lava composition and simulated volcanic production over the evolution of Mars, outgassing histories have been modelled \citep[i.e.][]{gillmann2009,gillmann2011,grott2011,leblanc2012,morschhauser2011,sandu2012,hu2015}. 

It is likely that water loss on Mars was stronger early in the evolution of the planet, both for hydrodynamic and non-thermal escape \citep{lammer2018}, and that for most of the evolution, escape was limited to a fraction of a bar. Some simulations indicate that ancient surface pressures on Mars, around 3.5-4 Ga, may have been low (a few to a few hundred millibar) and that much of present-day CO$_2$ atmosphere could originate from late outgassing \citep{gillmann2009,gillmann2011,leblanc2012,lammer2013}. Volcanic outgassing could also account for most of present-day water surface inventory (up to a 100 m global equivalent layer, \citet{alsaeed2019}). This is in line with work by \cite{Kite2014} suggesting an upper limit of $\sim$1 bar atmospheric pressure $\sim$3.6 Ga ago. This thin atmosphere was more prone to atmosphere loss than Earth's and Venus', accordingly over time favourable conditions for the presence of liquid water at the martian surface ceased.

\subsection{Venus}\label{sec:venus}

\subsubsection{Potential evolution paths} \label{sub:vpaths}
Compared to Earth and Mars, there is little information on the history of Venus. This is, in a large part, due to the overall young apparent age of most of the planetary surface, which limits observations regarding most of its evolution. No definitive evidence for surface water exists in the case of Venus, due to the lack of surface features dating back more than 300-1000 Ma. Surface mineralogy of tesserae (8\% of the present-day venusian crust) has been proposed to indicate the presence of felsic material in unknown quantities \citep{shellnutt2019}. This would suggest that an uncertain amount of liquid water could have existed on Venus for an unknown period of time. Most of the constraints on Venus' past are derived from its atmosphere that retains the accumulated effects of every process affecting volatile inventories. Needless to say, unraveling plausible scenarios is a complicated endeavour. Current research has not yet managed to converge toward a single agreed succession of events that would lead to present-day Venus.

If Venus and Earth formed in the same way from the same building blocks, the apparent dryness of the Venus could have been caused by its proximity to the Sun. Venus receives about 1.9 times more solar flux than Earth, and so it can relatively easily enter a runaway greenhouse state \citep{zahnle2007emergence}. However, while the atmosphere and surface of Venus are dry compared to Earth, the presence of water inside Venus has not been ruled out. Three different scenarios have been proposed to explain the lack of water at the surface of Venus \citep{Kasting88Icarus,Hamano+13Nature,gillmann2022}: (i) Venus was once an ocean planet like Earth, and then, its climate was destabilized by internal factors, and/or, as the Sun got brighter over time, the surface water was evaporated and lost by photolysis and escape of hydrogen. (ii) Venus evolved into a desiccated planet as it is today during its magma ocean crystallization process. (iii) Water was never outgassed from the interior of the planet and was trapped in its interior.

One-dimensional cloud-free climate calculations yield the radiation limit of about 280 W/m$^2$ considering a planetary albedo of about 0.2 for thick steam-dominated atmospheres \citep{Kopparapu+13ApJ,Goldblatt+13NatGeo}. Simply substituting these values into equation \eqref{eq:acr} gives the critical orbital distance of 0.82 au, seemingly suggesting that Venus was well inside the critical distance and thus followed scenario (ii). However, planetary albedo plays a critical role in defining the radiative balance of rocky planets. 1D climate models cannot fully handle the effects of atmospheric circulation, such as heat transport by winds and the distribution of relative humidity and clouds, which may favor the formation of an ocean. Recent 3D simulations have shown that the stability of water oceans on early Venus would depend on starting conditions in its atmosphere. If all water on early Venus was initially present as a liquid ocean in a moderate climate, clouds formed on the dayside would have reflected sunlight efficiently and maintained the surface water ocean considering a slow rotation rate of Venus \citep{Yang+13ApJL,Way+16GRL}. By contrast, if early Venus started out with a hot steam atmosphere, it would have remained in a runaway greenhouse state due to thermal blanketing by clouds formed on the nightside and strong absorption of sunlight by abundant water vapor \citep{Turbet+21Nature}. If Venus experienced a giant impact at the end of its accretion, like Earth, the latter case seems more plausible. On the other hand, there is a possibility that Venus had a different initial thermal state when it formed in a much different way from Earth \citep[e.g.][]{Jacobson2017}. This topic is out of scope of this chapter.

In scenarios (i) and (ii), water vapor is considered to be dissociated and lost by extensive escape of hydrogen to space. If Venus started with a water equivalent of one Earth' ocean mass, a preferential escape of hydrogen might have left about 240 bars of oxygen on Venus. Some fraction of oxygen may also have escaped directly if the energy input is high enough (possible roughly before 4 Ga). However, based on recent numerical results \citep{Johnstone20ApJ,Johnstone+21EPSL,saxena2019sun}, the activity of the young Sun was not strong enough to drive rapid oxygen escape, thus some oxygen would have remained in the Venusian atmosphere. 


The fate of the remnant oxygen has been considered to be a constraint for the recent evolution of the Venusian atmosphere. If oxygen builds up, the mixing ratio of water vapor drops so that oxygen could become less prone to escape \citep{Kasting&Pollack83Icarus,Wordsworth&Pierrehumbert14ApJ}. It had long been suggested that a mechanism that could explain the loss of oxygen from the atmosphere of Venus (and other terrestrial planets) was long-term ongoing loss of volatiles by non-thermal escape processes (see for example \citet{gronoff2020atmospheric}). However, total loss of O through non-thermal escape processes is estimated to be far less than those needed to explain the loss of an amount of water comparable to one Earth ocean \citep{gillmann2020,Persson+20JGR}. Rapid water loss requires additional oxygen sink(s) other than thermal and non-thermal escape processes, which can efficiently dispose of the remaining oxygen.

\vspace{5mm}

In scenario (i), some of the oxygen may have reacted with, for example, CO to form CO$_2$ (or reduced species outgassed at the time) and some of the oxygen may have been consumed by oxidation of Venusian crustal material (experiments by \citet{Berger2019} and \citet{filiberto2020}, models by \citet{gillmann2020,warren2023}, and review by \citet{gillmann2022}). However, it is unclear whether a combination of all these processes could have removed tens to hundreds of bars of O$_2$ immediately without preventing hydrogen from escaping. Generally speaking, current results suggest that only a moderate amount of oxygen can be extracted from the atmosphere by solid-gas reactions. Experiments and models currently suggest the total removal of O$_2$ may be limited to amounts similar to those caused by non-thermal escape (about a fraction of a bar to 1 bar at most). This does not outright make scenario (i) implausible but rather limits the possible range of water inventories (perhaps to a fraction of an Earth ocean, or arid conditions) and the time/length of the "wet" period (during the first few hundred million years). Recent studies suggest that a runaway greenhouse transition with enough water in the atmosphere (a minimum of a few tens of bars) may have caused global melting of the surface, which may become a volatile and oxygen sink \citep{gillmann2022,warren2023}. However, no evidence of such an event has been recorded due to the young age of the surface of Venus. This remains an active research topic, but the lack of constraints and multiple processes involved still make quantitative estimates unreliable. 

\vspace{5mm}

In scenario (ii) the remnant oxygen would have been immediately removed from the atmosphere by oxidizing ferrous iron in the surface magma e.g. \citep{gillmann2009b,lammer2018}. Compared with the solid rocky surface, the surface magma over the vigorously convective magma ocean is expected to react with the overlying atmosphere more efficiently, due to its higher mobility and resurfacing rate. Furthermore, modeling results suggest that, in scenario (ii), the planetary surface can remain molten until most of its water is lost \citep{Hamano+13Nature}. This is because the surface temperature could exceed the solidus temperature ($\sim$~1,340~K for peridotites) as far as a few tens of bars of steam atmosphere remains \citep{Hamano20OUP}. On the other hand, a recent numerical study \citep[][]{selsis2023} suggest that, under some conditions, a water-vapor atmosphere may see its structure dominated by radiative rather than convective layers, leading to cooler surface temperatures than for an adiabatic temperature profile. As a result, the occurrence and/or lifetime of a magma ocean may be greatly reduced. The magma ocean could solidify earlier than previously assumed, before the steam atmosphere escapes, which would reduce its efficiency as a trap for volatile species, including oxygen. 

Nonetheless, if the magma ocean acts as a volatile sink, the oxygen buildup associated with hydrodynamic escape would not have occurred when Venus followed scenario (ii). One of the implications expected from this scenario is that the Venus' magma ocean may have a redox evolution different than Earth's, depending on the amount of the remnant oxygen reacted. The oxidation of the magma ocean associated with extensive loss of water may affect the speciation and partitioning of volatile elements. It may also explain an elevated atmospheric N inventory on Venus, compared with Earth \citep{Wordsworth2016EPSL}. Due to the lack of direct samples, the redox state of the present Venusian mantle remains poorly constrained. As deduced from FeO contents of surfacial basalts, the overall oxygen budget may be similar to Earth's \citep[e.g.][]{Righter+2016AM}, whereas thermodynamic calculations and spectroscopic data suggests that the oxygen fugacity of Venus' surface may be as high as the magnetite--hematite phase boundary \citep{Fegley+1997Icarus}.

\vspace{5mm}

Accumulation of oxygen is not an issue in scenario (iii) either, as water would have only be a minor component of early outgassing in the first place. Venus surface and atmosphere would be dry, but its interior may not be. The potentially different dryness of the interior among the three scenarios might affect the subsequent evolution of the interior. In such an evolution, inefficient outgassing of the magma ocean \citep{ikoma2018,solomatova2021,Salvador&Samuel2023} combined with high surface pressures \citep[increasing solubility of water in the magma ocean][]{gaillard2014}, would have limited outgassing of water. Early loss of hydrogen could have favored the oxidation of the magma ocean \citep{Gaillard2022}, therefore an increased loss of H on Venus may help explain the divergent evolution path of Venus. Later water outgassing related to volcanism would be prevented by high surface pressures due to the thick CO$_2$ atmosphere \citep{gaillard2014}. 

\subsubsection{The interior of Venus} \label{sec:venus_interior}

Our understanding of the interior of Venus is very limited. Observations are generally consistent with an approximately Earth-like internal structure and composition, but error bars are large and accommodate a wide range of possible configurations. In short, no direct observations strongly imply any radical structural differences compared to Earth, despite the surface expressions of interior dynamics diverging widely from what we can observe on Earth.

It is consensus that Venus displays no present-day global plate tectonics, but is possibly (1) in a plutonic squishy lid regime because of its high surface temperature or (2) in a quiet period of an episodic lid regime, or (3) in stagnant-lid after a transition from mobile-lid regime. Specific sequences of regime transitions have been discussed by \citet{Rolf2022} and \citet{gillmann2022} (and references therein), but so far the question remains unsolved. The absence of a global self-generated magnetic field on Venus \citep{nimmo2002does,smrekar2018venus}, possibly caused by insufficient heat extraction out of the core, is likely tied to the question of the mantle dynamics evolution through time \citep{ORourke2018}. Further investigation into past evidence of magnetic field generation may provide new constraints for the evolution of Venus' mantle thermal state and convection regime \citep{o2019detectability}. 

Differences in the dynamics of the interior of Earth and Venus and convection regimes may be related to the suspected dryness of the interior of Venus
\citep{kiefer1986,grinspoon1993implications}. Because water affects melting and the mantle viscosity,it also affects mantle dynamics: for example, a dry mantle produces less melt and has a higher viscosity; \citet{green2014,karato2015water}. However, the extent to which water can be a critical factor still remains to be fully analysed and quantified. Additionally, some works suggesting dry Venus \citep{kiefer1986,grinspoon1993implications} made strong assumptions on the processes at work on Venus. \citet{kiefer1986}, for example, linked the correlation between Venus' geoid and topography to the absence of an asthenosphere, which could be due to dry conditions (and is now questioned; \citet{armann2012,maia2022lithospheric}). The dryness of the interior inferred by \citet{grinspoon1993implications} relied on assumptions on the rate of volcanic production and outgassing. However, volcanic production is poorly constrained and highly debated \citep{herrick2023}, while outgassing itself would be affected by Venus' high surface pressure and reduced water release (see Section \ref{sub:volatiles}). As a result, whether Venus' interior is dry or wet remains uncertain.

So far, no measurement has allowed us to rule out a specific set of evolution scenarios with any degree of certainty. We have no sample of volcanic rocks or volatiles from Venus, to estimate its interior composition and conditions. Data on the state of the core are consistent with a solid, liquid or stably stratified core \citep{dumoulin2017}. Models \citep{ORourke2018} suggest that if Venus had an Earth-like structure and composition, it would likely generate a similar global magnetic field. Therefore, without a better understanding of the interior of the planet, we are unable to decide whether Venus lost its global magnetic field or never generated one in the first place. 

This need for more observations is driving the current Venus exploration projects with multiple missions planned or selected: DaVinci (USA), VERITAS (USA), EnVision (Europe), Venus Orbiter Mission (also known as Shukryaan-1) (India), Venera-D (Russia) and VOICE (China). Those missions are set to improve our knowledge of the interior structure of the planet, state of its core, its subsurface structure, surface topography, geology and composition and atmosphere structure and composition (both close to the ground and higher up, with a special interest in volcanic production), with a focus on traces of activity through time.

\subsection{Earth} \label{sec:earth}

\subsubsection{Oxidation of Earth's magma ocean}
Compared with other planetary bodies such as Mars and the Moon, Earth has a highly oxidized mantle characterized by high Fe$^{3+}$ abundance \citep[e.g.][]{Frost+2008}. The evolution of the redox state of the Earth's mantle is one of the research fields actively explored (chapter 00008). Geochemical studies of Hadean zircons suggest that the oxidation of the Earth's mantle might have occurred very early \citep{Trail+2011Nature}, possibly during its magma ocean stage.

One of the mechanisms proposed to explain the oxidation of the terrestrial magma ocean is hydrogen escape accompanying dissociation of water \citep{Hamano+13Nature,Sharp+2013EPSL}. The present Earth's upper mantle has a Fe$^{3+}$/$\Sigma$Fe of 0.02--0.06, which corresponds to the Fe$_2$O$_3$ content of 0.16--0.48~wt\% \citep[][and references therein]{Hirschmann22GCA}. Using the simple assumption of oxidation of FeO to Fe$_2$O$_3$, the amount of water required to raise the Fe$_2$O$_3$ content from zero to the present level is equivalent to 0.5-1.5 Earth's ocean masses for the whole mantle, and about 1/3 of this value (i.e. 0.2-0.5 ocean masses) for the upper mantle only. Earth is sufficiently far from the critical orbital distance \eqref{eq:acr} discussed in Section \ref{sec:MO-cooling}, thus Earth's magma ocean would have crystallized rapidly on the order of 1 Myr or less \citep{Hamano+13Nature, Lebrun+13JGR, Salvador+17JGR, Bower+22PlanetSciJ}. \citet{Hamano+13Nature} estimated the total loss of water throughout the Earth's magma ocean stage to be about 0.2 ocean masses at most for a pure steam atmosphere. Due to the low mixing ratio of H-bearing species in the upper atmosphere, the loss of water would become probably lower if the early atmosphere included large amounts of other gaseous species such as CO$_2$. Thus, water loss during the Earth's magma ocean stage might have a partial contribution to oxidation of the upper mantle, but the effect could be marginal for the oxidation of the whole mantle. 

An alternative mechanism is the redox disproportionation of Fe$^{2+}$ to Fe$^{3+}$ and Fe$^0$. This oxidation mechanism was originally proposed to be driven by crystallization of a solid phase with affinity for Fe$^{3+}$, such as perovskite. However, it is unclear how quickly Fe metal formed by iron disproportionation could sink to the core through the solid mantle. As described in section \ref{sec:MO-dispro}, recent experimental and theoretical studies reported that redox disproportionation could occur directly in a magma ocean owing to a high stability of Fe$^{3+}$ at high pressure \citep{Armstrong+19Sci,Deng+20NatCom,Kuwahara+2023NatGeO}. Metal silicate equilibration at the base of the magma ocean can produce a high Fe$^{3+}$/$\Sigma$Fe in the melt even at an oxygen fugacity as low as IW-2. This oxidation mechanism requires that the base of the magma ocean attains pressures as high as tens of GPa. This is in line with estimations of terrestrial magma ocean depths caused by giant impacts during the late stage of Earth's formation \citep{Nakajima2021}. The Earth's large planetary mass and the accretion history involving giant impacts during the late stage of formation may be key factors that set the different initial condition for the redox evolution of its mantle.

This oxidation scenario seems promising for Earth, although the estimated Fe$^{3+}$/$\Sigma$Fe in the silicate melt shows significant differences among studies. Based on the assumption of metal-silicate equilibration at 25-90~GPa, \citet{Deng+20NatCom} computed the Fe$^{3+}$/$\Sigma$Fe of about 0.01-0.02, which is comparable to the value observed in the Earth's upper mantle. On the other hand, the Fe$^{3+}$/$\Sigma$Fe estimated by \citet{Armstrong+19Sci} and \citet{Kuwahara+2023NatGeO} are much higher. Using a relatively low base pressure of 25-33~GPa, \citet{Kuwahara+2023NatGeO} obtained a Fe$^{3+}$/$\Sigma$Fe value of 0.2-0.5, which is one order of magnitude higher than the Fe$^{3+}$/$\Sigma$Fe in the upper mantle. The cause of such differences remains unclear. If the excess in the Fe$^{3+}$/$\Sigma$Fe is confirmed, then it may suggest reduction processes such as the late accretion of reducing material \citep{Dauphas17,F-G&Klein17,Genda+17}, which could form a transient reducing surface environment, or may reflect incomplete metal-silicate equilibration, rather than a single equilibration event at the base of the magma ocean \citep[e.g.][]{Rubie+15TreatiseGeophys}.

\subsubsection{Earth's most distinctive feature: Life affects planetary evolution and volatile cycles}

The discussion above has made abundantly clear the fact that the Earth today is very different to present-day Venus and Mars. The present-day Earth is characterised by an equable climate (from the point of view of life), stable water at its surface, an oxidised atmosphere, and a robust plate tectonic regime that links surface processes to those in the mantle, in the process buffering the whole system Earth. 

At present, Venus’ surface is dry and hot with a less oxidized atmosphere (and H$_2$SO$_4$ clouds), and no plate tectonics. Mars is a dry cold planet with a CO$_2$ atmosphere and a stagnant lid. Neither has life at its surface and the question of whether Venus could host some form of life in its clouds, or Mars in its subsurface, remains debated.

However, as the above discussions point out, the Earth did not start out as it is today. Indeed, life could not have emerged on an oxidised planet since abiogenesis (the emergence of life, as we know it) requires reduced organic building bricks (e.g. \cite{hudson2020co2}. On the other hand, the oxidation that is so characteristic of its surface is a product of biological processes. Thus, while the early planets, Earth, Venus and Mars may have shared certain characteristics that permitted life to emerge, their evolution paths diverged significantly. As we understand the emergence of life, it was conceivable on Mars but remains uncertain for Venus, due to the remaining uncertainties regarding early surface conditions (see section \ref{sub:vpaths}). Since life is an integral component of the Earth system, it is pertinent here to consider, in the context of terrestrial geology and geophysics, how it appeared and evolved through geological time. Since reconstructions of the past Earth's atmosphere composition do not highlight significant changes in bulk composition related purely to long-term escape \citep{gronoff2020atmospheric}, the intimate interaction between the effects of life at the surface, on internal processes, as well as on the atmospheric envelope and vice versa, is of equal importance. 

There are numerous theories about how and where life emerged, in deep ocean hydrothermal vents \citep{baross1985submarine, russell1997emergence,martin2003origins}, shallow water hydrothermal vents \citep{westall2018hydrothermal}, subaerial hydrothermal vents \citep{damer2020hot, van2021elements}, in impact craters \citep{sasselov2020origin}, in deep-seated fault systems \citep{schreiber2012hypothesis}, in pumice rafts \citep{brasier2011pumice} or in nuclear sands \citep{maruyama2023nuclear} to name but a few hypotheses (for a review see \citep{westall2018hydrothermal}). \cite{Westall2023} briefly reviewed the conditions that existed during the Hadean (4.5-4.0 Ga), and that were apparently compatible with prebiotic processes and the emergence of life. They included an anoxic atmosphere, (i) a high EUV flux reaching the surface of the Earth that may or may not have been beneficial for early prebiotic processes, depending upon the scenario preferred, (ii) a highly volcanic environment including abundant hydrothermal activity, and (iii) a largely ocean-covered planet with little exposed land mass. Modern-style plate tectonics did not exist and there is still much debate as to what style of tectonic regime was active during the Hadean \citep{capitanio2020thermochemical,gerya2022numerical}. A Hadean initiation of plate tectonics has been hypothesised on the basis of an andesitic source for zircons dating back to the Hadean \citep{turner2020andesitic}, while evidence for both vertical and horizontal deformation styles has been documented in Paleo-Meso Archean formations \citep{heubeck2023malolotsha}, suggesting some form of tectonic activity. Recent studies suggest that modern style plate tectonics did not arise until much later in Earth's history, towards the end of the Proterozoic, about 800 Ma (see Figure 7, \citep{Hawkesworth2020,stern2018evolution}, therefore the plutonic squishy lid might offer an alternative explanation \citep{Rozel2017,Lourenco2018}.

\begin{figure}[tb!]
\centering
\includegraphics[width=0.8\linewidth]{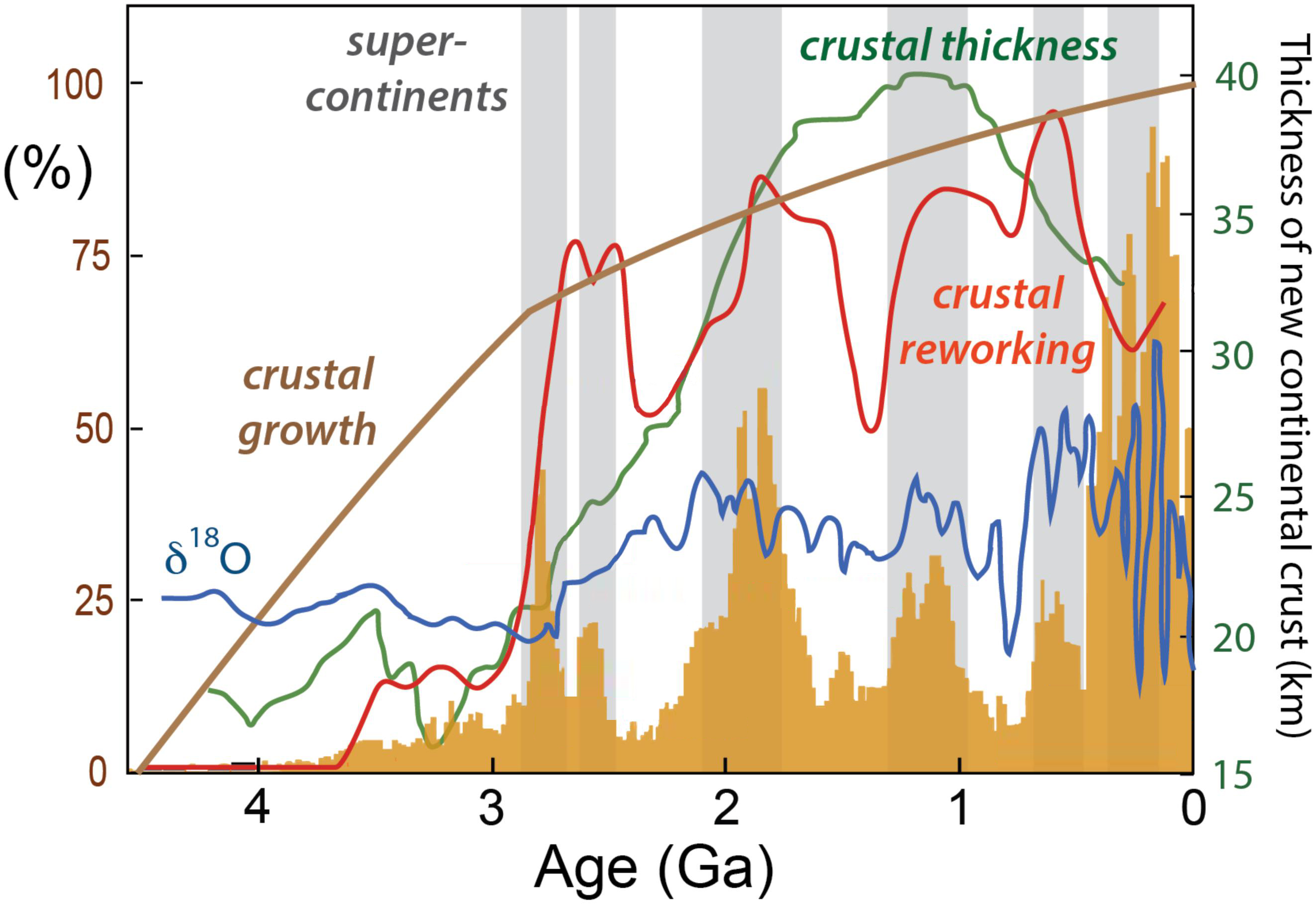}

\caption{Indicators of crustal evolution and the gradual rise of modern style plate tectonics, using zircon crystallization age distribution (lower serrated yellow curves), throughout geological history. Vertical grey bars mark the development of supercontinents/supercratons (see references in \citet{Hawkesworth2020}). The crustal growth curve (brown) is adapted from modelling by \citet{dhuime2012change}. The crustal reworking curve is based on the Hf isotope ratios of zircons in different time slices \citep{dhuime2012change}. Variations in the thickness of continental crust at the sites of crust generation through time is inferred from Rb/Sr measurements. The variation of $\delta$$^{18}$O values in zircons also marks the shifts in the composition of juvenile crust away from mafic compositions \citep{Hawkesworth2020} (adapted from \cite{Hawkesworth2020}).}
\label{fig:Picture1_FW}
\end{figure}

As soon as it appeared, maybe multiple times before really gaining a foothold, early life forms were intimately linked to the volcanic substrates from which they emerged. Note that all the proposed scenarios for the the emergence of life referred to above were located in volcanic settings. They were chemolithotrophic microorganisms obtaining their energy from oxidation of inorganic materials, such as H$_2$, or through redox alteration of reactive minerals at the surface of the volcanic rocks (e.g. olivine, pyroxene, metal sulphides), or chemoorganotrophs oxidising organic substrates, such as dead organisms or bio-available abiotic organics \citep{Weiss2016}. At this stage, life was completely dependent upon the mineral substrates. At some stage, through mutation, life developed the ability to capture energy from sunlight (phototrophs, \citep{martin2018physiological}), thus freeing it of the mineral substrates and providing it with access to an unlimited source of energy. The oldest, well-preserved terrestrial rocks dating back to about 3.5 Ga testify to the already widespread presence of anoxigenic phototrophs in shallow water environments. 

\begin{figure}[tb!]
\centering
\includegraphics[width=0.8\linewidth]{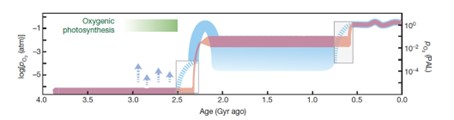}
\caption{Relative concentrations of atmospheric oxygen throughout Earth’s history, including the classical two-step evolution (orange line, Kump, 2008) and the more recent model including unstable climate and O$_2$ concentrations after the first rise of O$_2$ concentrations \citep[][]{lyons2014}. Possible “whiffs of oxygen” (blue arrows) during the Neoarchean -- and even before according to \cite[][]{Anbar2007} (adapted from \citep{lyons2014})}
\label{fig:Picture1}
\end{figure}

While access to unlimited sunlight gave the phototrophs an energy-rich, metabolic advantage over the chemotrophs (litho- and organ-) enabling them to grow faster and create greater biomass, it was the mutation that led to oxygenic photosynthesis that would change the Earth, both physically, chemically, and biologically. The timing of this biological invention is debated with certain geochemical studies suggesting that “whiffs of oxygen”, i.e. biological oxygen, may have occurred as early as the Paleo-Mesoarchean ($\sim$3.5-3.0 Ga) and during the next 500 Myr (see Fig. \ref{fig:Picture1}; \citet{Anbar2007}). Certainly by 2.9-3.0 Ga there is geochemical evidence for free oxygen and bioconstructions, such as stromatolites, start becoming common. It has been suggested that changes in the redox state of the Earth’s surface during the NeoArchean-Proterozoic transition may have been influenced not only by the production of oxygen by photosynthesisers, but also by changes in the redox state of the mantle, that affected the redox of outgassing species from volcanic eruptions (\citet{wang2020mantle}), although others have suggested that there was little change in the oxygen fugacity of the Archean mantle (e.g. \citet{nicklas2018redox}). Whatever the major influences during the Neoarchean, during this period, huge formations of stromatolites produced by oxygenic phototrophs occur in Neoarchean formations across the world and, indeed, dominated the paleontological record for most of the Proterozoic. 

The turning point was when the oxygen contents in the atmosphere started slowly to rise between about 2.4--2.3 Ga (the so-called “Great Oxygenation Event” (GOE), \cite{Holland2002}, see Fig. \ref{fig:Picture1}). In fact, the gradual accumulation of oxygen depended upon a delicate interplay of sinks and sources that probably started as early as about 3.0 Ga \citep{lyons2014} and may have continued until about 800 Ma \citep{lyons2021oxygenation}. As noted above, we keep in mind, however, that there may not have been major changes to the redox state of the mantle \citep{nicklas2018redox}. The production of oxygen was balanced by its consumption by the initially totally reduced surface of the Earth, including reduced organic carbon, sulphide minerals, other ferrous minerals, and reduced volcanic gases, such as H$_2$S, as well as biogenic methane and its photodissociated product ethane. Concurrent with the gradual oxidation of the Earth’s surface envelope, including its atmosphere, gradual changes in mantle buffering resulted in changes in the composition of volcanic gases, more reduced during the Archean, but more oxidised after the GOE. 

While the presence of oxygen in the atmosphere led to elimination of methane as greenhouse gas, there is still the problem of reducing CO$_2$ in the Neoarchean atmosphere. The isotopic ratios of carbon in the Neoarchean are relatively constant, suggesting that there was not a major change in the burial flux of the total carbon inventory. However, one solution to the conundrum could be that the flux of buried organic carbon during the Archean may have been far higher than previously believed \citep{krissansen2021carbon}. Whatever the processes involved, oxygenation of the atmosphere had the catastrophic consequence of lowering surface temperatures because the Sun’s luminosity was still about 20-30\% lower than today (as was the case during the Hadean and Archean, \cite{Sagan1972}), which led to the first series of global glaciations, the Huronian glaciations between 2.45 and 2.22 Ga, when low latitude continental terranes were covered with glaciers \citep{bekker2012oxygen}. This effectively cut off phototrophic production of oxygen, increased temperatures due to the accumulation of volcanic CO$_2$ that led to melting of the glaciers and a sudden massive influx of nutrients that fed the rapid re-development of the oxygen producing cyanobacteria and the precipitation of biomediated carbonate deposits, the so-called cap carbonates, \cite{Hoffman1998}. The sudden explosion of biomass production led to a very large rise in oxygen levels between 2.3 and 2.06 Ga, called the Lomagundi-Jatuli event, during which oxygen levels almost reached present-day levels \citep{Melezhik2013}, before falling to very low atmospheric partial pressure levels of O$_2$ < 0.1\% of present atmospheric levels \citep{lyons2014} that reigned for the rest of the Proterozoic until about 0.8 Ga \citep{Canfield1998}, as illustrated by Figure \ref{fig:Picture1}. It is thought that this long period of stasis during the Meso-Neoproterozoic was due to the fact that only the upper part of the oceans were oxygenated while the deep ocean remained reduced. 

However, it was during the Proterozoic that the extra energy available via oxygenic photosynthesis fuelled the mutations giving rise to eukaryotes, i.e. organisms with a membrane bound nucleus (prokaryotes, including Bacteria and Archaea do not have membrane bound nuclei). Indeed, it was only after complete oxygenation of the entire water column and the stabilisation of nutrient cycles and environmental conditions could life evolve into the diverse biota that characterised the terminal stages of the Neoproterozoic (1-0.54 Ga) and the dawn of the Cambrian period \citep[][]{Planavsky2015}. The Neoproterozoic period was characterized by extreme biogeochemical and climatic volatility that led to a series of very severe, low latitude glaciations \citep[][]{Hoffman1998,Canfield2005,lyons2014} resulting in a highly unstable carbon cycle. This period of multiple “snowball Earths” could have been triggered or at least influenced by the onset of modern plate tectonics \citep[e.g.][]{stern2018}. The global distribution of continents and their positions relative to the equator likely also had an effect on albedo and therefore temperatures \citep[][]{Endal1982,Schrag2002}. At the same time, changes in oceanic ecosystems led to the rise of eukaryotic algal primary producers as opposed to the prokaryotic cyanobacteria so prevalent from the Neoarchean (3.0-2.5 Ga) through the Mesoproterozoic (2.0-1.0 Ga), and the eventual emergence of metazoan life. Add to this the rise in oxygen globally and the conditions for significant evolutionary changes to life were all present \citep[][]{Planavsky2015}. 

Today, there appears to be a delicate balance between atmospheric composition, especially the greenhouse gases CO$_2$ and CH$_4$, the presence of water at Earth's surface, subduction of organic matter and carbonates, and subsequent release of volcanic CO$_2$ back into the Earth's atmosphere (see \citet{charnay2020faint} and \citet{catling2020} for recent reviews). It was likely established gradually throughout the geological evolution of the Earth, possibly as far back as the Paleoarchean or even before during the Hadean. The exact process for the recycling of this early carbon cycle is still unknown, but it was most likely different from modern-day plate tectonics. Initially, the higher CO$_2$ concentration in the Hadean/Archean atmosphere would have offset the lower luminosity from the Sun \citep[i.e.][]{Kasting88Icarus}. As solar luminosity increased, CO$_2$ was gradually removed from the atmosphere by the silicate/carbonate weathering cycle and sequestration of carbonates and organic carbon into the crust. In this way, throughout time, climatic stability has been ensured -- as has been the presence of liquid water, one of the prime necessities of life on Earth.


\section{Conclusions}
\label{sec:conc}

The range of possible terrestrial planet evolution is difficult to explore with only three very particular examples, however, even in our Solar System, the diversity among the rocky planets is striking. The exact causes for the clear differences between present-day Mars, Venus and Earth are still uncertain. It is now suspected that both their early (first few 100 Myr) and long-term evolution had a role to play. Atmospheric and interior conditions are the result of billions of years of accumulated and interacting processes that affect the physical and chemical properties of every planetary layer. 

In recent years, our understanding of how magma ocean processes affect the early evolution of terrestrial planets has grown considerably. 
Studies show that in case the terrestrial planets started similarly in a hot molten state, the iron disproportionation at high pressures in a magma ocean can set a different initial condition for their subsequent redox evolution. Based on the assumption that metal-silicate equilibration occurred at the base of the magma ocean, due to their higher mass, Earth and Venus could have had a deep magma ocean with a higher abundance of Fe$^{3+}$ and an oxidizing atmosphere consisting of CO$_2$, N$_2$ and SO$_2$, while a shallow magma ocean on Mars would be relatively Fe$^{3+}$-poor and overlaid by a weakly reducing atmosphere consisting of CO, CO$_2$, and N$_2$. On all the three planets, the majority of H would be dissolved in the magma ocean as H$_2$O at an early stage, and then would have outgassed at the surface as solidification proceeded. 

In the case that a steam-dominated atmosphere forms, the subsequent evolution of the terrestrial planets depends on their orbital distance from the Sun. Earth and Mars could be sufficiently far from the Sun to keep their water intact against hydrodynamic escape owing to their short crystallization period. On the other hand, for Venus, three main different scenarios are possible: (i) rapid crystallization where water is retained or (ii) desiccation during a long-lived magma ocean period or (iii) that a high pressure CO$_2$ dominated atmosphere suppresses water outgassing. In the second scenario, the Venusian magma ocean can be oxidized by remnant oxygen produced by photodissociation of water, followed by preferential escape of hydrogen. At this point, the redox evolution of Earth and Venus may also diverge. If H$_2$O outgassing is insufficient to form a steam-dominated atmosphere, such an evolutionary dichotomy would not appear, and early planetary mantles would contain high abundance of water as discussed in scenario (iii). Planetary evolution during a magma ocean phase would set the initial conditions in terms of redox state in the mantle and atmosphere. This can have consequences for the crystallization history, the total and interior inventory of water and the climate throughout the entire planetary evolution.

Tectonic regimes can be divided into three main types of convection regimes: plate tectonics, plutonic-squishy lid, and stagnant lid. Present-day Earth lies in the plate tectonics regime and features high heat flow and volcanic production rates. The onset time of modern plate tectonics on Earth is still under debate and might be related to the presence of life on Earth. The onset of plate tectonics depends on the rheology of the planetary lithosphere and the stresses in the convective mantle. Although surface temperature and water abundance are considered to be important factors affecting the rheology, whether they favor plate tectonics or not is still debated due to complicated feedbacks. The presence of subduction zones favors volatile exchanges between the atmosphere and the mantle. Mars has probably been in the stagnant lid regime for billion of years with limited geological activity in the last billion years. The heat flux in the stagnant lid regime is expected to be low, and this could explain why its core is still in a mostly liquid state, while not featuring a present-day core dynamo. Its smaller mass and more reduced interior is expected to limit total outgassing from its interior. 

In contrast to Earth and Mars, the tectonic regime of past and present Venus remains unclear. Present-day Venus does not display signs of Earth-like plate tectonics, but seems to be more active than Mars and its stagnant lid regime. Its relatively young surface obscures most signs of prior tectonic and volcanic activity. A major constraint on Venus' interior is that at present it does not feature a global magnetic field. As Venus and Earth are comparable in size and mass, their differences may originate early on during their respective accretion and magma ocean stage.  

In addition to volatile exchanges between surface and the interior, atmospheric escape is also a crucial process shaping and driving the evolution of planetary atmospheres. During the early evolution, hydrodynamic escape and impact erosion are important processes. Hydrodynamic escape is driven by strong EUV flux from the young Sun or by heating from a hot surface. Intense hydrodynamic escape would occur when the upper atmosphere is rich in H-bearing gaseous species such as H$_2$ and H$_2$O. This process would efficiently remove any primordial H$_2$-He atmosphere the planetary embryos obtained from the solar nebula during their formation. Also this process would control the crystallization rate of a magma ocean located under a runaway greenhouse steam atmosphere. Impact erosion removes portions of planetary atmospheres. The most efficient loss occurs on small planets like Mars and/or in the presence of a surface water ocean. Impacts can also supply material, including volatiles. The net effect largely depends on the volatile content of the impactors, but also on the specific impact conditions. 

As both impact and EUV flux decrease with time, non-thermal escape processes gains importance. Non-thermal escape involves photochemistry and charged particle interactions with the solar wind and among particles. Particles that obtained sufficient energy from the interaction can escape. Small planets like Mars are again more vulnerable to loss of heavier volatile species, such as O, C, N, and noble gases. Water loss, through escape of H and O ions is possible on all three planets at a comparable, but relatively slow rate. Based on recent measurements the previous idea that magnetic fields would provide an efficient shield against non-thermal escape is now being questioned as escape rates of heavy ions are comparable on Earth (featuring a global magnetic field), Venus, and Mars (no present-day core dynamo). 

Upcoming missions to Venus specifically aim to shed light onto its past evolution compared to those of Mars and Earth. Likewise, sample return missions from Mars should lead to a dramatically better understanding of both Mars interior and surface conditions, as well as whether Mars might be currently hosting life in the subsurface. The future characterization of a large number of terrestrial exoplanets will provide further reference points for the possible evolution paths of rocky planets at all stages of their histories, including the critical magma ocean phase. The combination terrestrial planet studies and exoplanet observations will provide both a detailed and statistic view of terrestrial evolution pathways and will help us to solve remaining mysteries that we have touched upon within the framework of this chapter.

\section*{Acknowledgements}
The authors thank the anonymous reviewer for his constructive comments that helped to improve the manuscript. K.H. acknowledges financial support by JSPS KAKENHI grant No. JP18K13603 and MEXT KAKENHI grant No. JP22H05150.
CG acknowledges that this work has been carried out within the framework of the NCCR PlanetS supported by the Swiss National Science Foundation under grants 51NF40$\_182901$ and 51NF40$\_205606$.



\bibliographystyle{elsarticle-harv} 
\bibliography{refs}





\end{document}